\documentclass[10pt, conference, letterpaper]{IEEEtran}
\usepackage{float}
\usepackage{amsmath}
\usepackage{amssymb}
\usepackage{amsthm}
\usepackage{tabularx}
\usepackage{graphicx}
\usepackage{subfigure}
\usepackage{authblk}
\usepackage[linesnumbered, ruled]{algorithm2e}

\usepackage{threeparttable}
\usepackage{slashbox}
\usepackage{algorithmic}
\usepackage{tabularx}
\usepackage{booktabs}
\usepackage{amsfonts}
\usepackage{dsfont}
\usepackage[all]{xy}

\def\squarebox#1{\hbox to #1{\hfill\vbox to #1{\vfill}}}

\begin{document}

%\title{A Cyber-Human Interaction Approach \\ on Indoor Localization}
\title{A Cyber-Human Interaction Based System\\ on Mobile Phone for Indoor Localization}
%\title{A Cyber-Human Interaction Based System on Mobile Phone for Indoor Location-Based Service}

%\author{}
\author{Zimu Yuan  \\ 
University of Chinese Academy of Sciences, China\\
Institute of Computing Technology, Chinese Academy of Sciences, China\\
zimu.yuan@gmail.com
}

\maketitle \thispagestyle{empty}

\begin{abstract}
In this article, we study the Cyber-Human Interaction (CHI) based approach that the "Human" part sets a list of location-based objectives and makes the pathway decision whereas the "Cyber" part provides the pathway suggestion, infer heuristics from the environment along the pathway and incrementally resolve the location-based objectives with new heuristics for indoor localization. For this study, we implement a CHI-based system on mobile phone. The CHI-based system offers the pathway suggestion and the solution of the location-based objectives based on its trajectory management. Without any priori knowledge on the area of interest and any aid from other equipments, a laborer can achieve his location-based objectives by walking through the area of interest and simultaneously online interacting with the CHI-based system installed in his phone. In evaluation, we conduct the experiments and show the advantage CHI in reducing the time cost and the expense cost for the laborer.
\end{abstract}

%\vspace{1ex}

%\begin{IEEEkeywords}
%Node Placement, Flow demands, Multi-Hop Wireless Network
%\end{IEEEkeywords}

%\vspace{2ex}

%-------------------------------------------------------------------------
\section{Introduction}
Indoor localization is one of the hottest research area on mobile and pervasive computing. The majority of previous indoor localization methods take Received Signal Strength (RSS) of WiFi signals, which can be easily obtained by inexpensive mobile device, as a metric for location calculation. The most popular type of methods on utilizing RSS for localization is the Fingerprinting-based approach that collects RSS distribution of WiFi signals in known locations to build a RSS-to-location database and infer the location of interest points by this pre-built database. However, the Fingerprinting-based approach requires labour-intensive work to sample the RSS distribution and establish the RSS-to-location mapping. A laborer may walk to each sampling point, gather the RSS of WiFi signals from the APs by the mobile device, measure the exact location of the sampling point and record the tuple --- location, identifier of the APs and corresponding RSS value of signals  --- to the database. A vast workload comes to the laborer when a large number of points are needed to be sampled with the reason such as the sampling area being too large, a higher localization accuracy required. Further, the collected RSS-to-location data cannot be used permanently after the database being initially established. Once a change happens in the space, e.g., a new AP is added or a existing AP is moved, the laborer would be required to take the repeated process of measurement to calibrate the RSS-to-location mapping.

Recently, Crowdsourcing-based methods are proposed as an alternative approach \cite{Unloc} or an improvement \cite{Zee} to the Fingerprinting-based approach. The Crowdsourcing-based approach collects the RSS distribution by the movement of users with mobile device, estimates the trajectories of the movement of users,  and thus maps the RSS values to the estimated location of points in these trajectories. Compared to the Fingerprinting-based approach, the Crowdsourcing-based approach reduces or eliminates the planned labour-intensive work, and instead establish the RSS-to-location mapping by the unplanned movement of users(, or to say, the normal movement of users, such as walking from the office area to the rest area). However, the Crowdsourcing-based approach requires the involvement of the crowd, and produces a large volume of duplicated trajectories that have overlapping segments between each other. Also, the Crowdsoucing-based approach is inadaptable when one cannot persuade the crowd, or it is meaningless to employ the crowd or consume too much time to walking around the area of interest, e.g., one just want to know the roughly relative position between APs through several rounds of walking around the area.

Different with the Fingerprinting-based approach and the Crowdsourcing-based approach, we investigate on another approach --- Cyber-Human Interaction (CHI) --- to reduce the time and expense cost for indoor localization. The Fingerprinting-based approach requires planned labour-intensive work to establish RSS-to-location mapping in database, the process of which totally depend on the decision making by the laborer, and typically the Fingerprinting-based localization system does not provide heuristics on the choice of walking pathway for laborer, while the Crowdsourcing-based approach totally avoids the decision choice made the laborer, instead infers the heuristics from the trajectories generated by the unplanned movement of the crowd, and establish RSS-to-location mapping based on the inferred heuristics. However, from our perspective of view, both of these two approaches consider the extreme case for localization, and ignore the benefits that can be obtained by combining the heuristics indicated from the trajectories and the decision made by the laborer. More, the localization objective made by the will of the demander is varied in different situations, e.g., to draw the floor plan for the area of interest, to track the trajectories of users, or to estimate the location of APs and take the APs as the beacon nodes for localization. Following the will of the demander requires the localization system accepting the human specified input and being flexible to adjust its localization objective according to the situation. This type of localization system neither merely depends on the fashion of fignerprinting type that decisions are totally made by the laborer nor depends on the fashion of crowdsourcing type that does not come close to the will of the demander, but relies on the fashion of the cyber system(, which indicates heuristics from collected trajectories,) and human(, who makes the decision,) interaction type. In our proposed CHI-based approach, (1) initially, the laborer can set a list of localization objectives, and the CHI-based system provides the suggestion of the walking pathway for the laborer according to the order of the localization objectives; (2) as the laborer walks through the area of interest, new trajectories and their corresponding RSS-to-location mapping data are being collected, the CHI-based system incrementally add these data into the consideration of pathway suggestion, and remove the objectives that have been completed; (3) simultaneously, the laborer can re-order or change the localization objectives.

The CHI-based system of our implementation is a lightweight system on mobile phone that provides a set of location-based services not merely relying on the measurement of the Inertial Measurement Unit (IMU) sensors equipped in phone but also utilizing the data collected from the environment to help improving their quality. In details, we make the following contributions:
\begin{itemize}
  \item Without any assumption of priori knowledge and any aid from other equipments, a laborer can carry the phone with CHI-based system installed to explore the area of interest.
  \item CHI-based system is designed to online process the location-based objectives set by the laborer, which is flexible and could be adjusted in situation according to the data collected from the environment, unlike the offline localization system or the online location-based system with a back-end server.
  \item Trajectory management is implemented in CHI-based system, providing the function of evaluating the quality of the laborer's trajectories and fusing the trajectories to construct a compound trajectory of high quality.
  \item The location-based services implemented in the CHI-based system offers the pathway suggestion and the solution of location-based objectives for the laborer.
  \item We experimentally evaluate the CHI-based approach and show its advantage in reducing the time cost and the expense cost for the laborer.
\end{itemize}

The rest of this article is organized as follows. Section \ref{overview} gives a overview of the CHI-based system. Section \ref{tra_mgm} presents the trajectory management of the CHI-based system. Section \ref{sec_lbs} introduces the location-based services implemented in the CHI-based system. Section \ref{use_case} describes a use case of the CHI-based system. Section \ref{evaluation} presents the evaluation results. Section \ref{related} summarizes the related works. Section \ref{conclusion} concludes this article.

\section{Overview} \label{overview}

\begin{figure}
\centering
\includegraphics[width=3.8in]{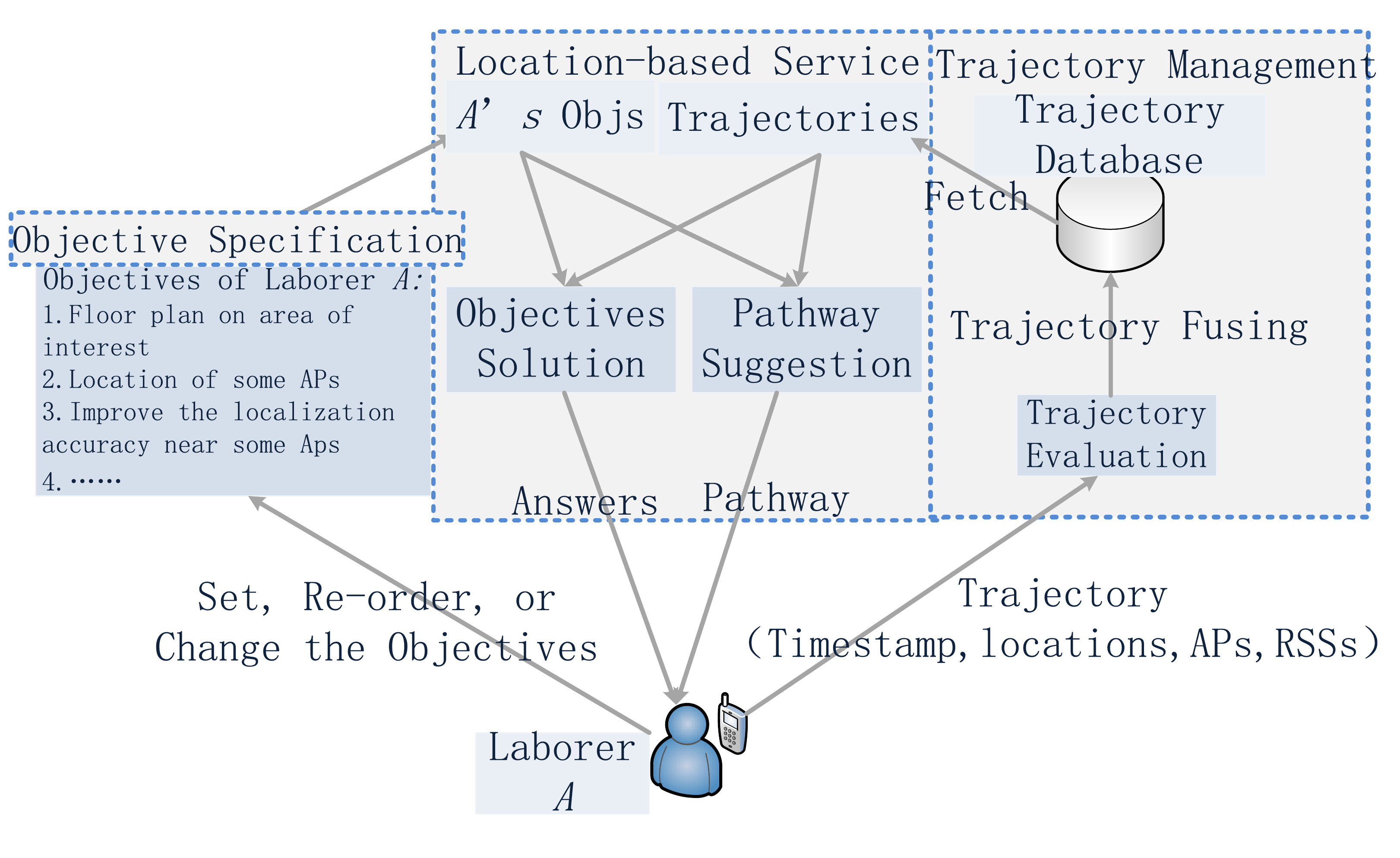}\\
\caption{\textrm{The framework of CHI-based system}} \label{fig1}
\end{figure}

The CHI-based system (hereafter called CHI for short) does not assume any priori knowledge on the area of interest. When the laborer set a list of location-based objectives, CHI provides the pathway suggestion for the laborer; When the laborer walking in the area of interest, CHI collects and measures the trajectory of the laborer by the data obtained from IMU sensors and by the process of AP-based calibration; When a new trajectory is recoded, CHI incrementally add it into its consideration of resolving the location-based objectives set by the laborer.

The framework of CHI is shown in Figure \ref{fig1}. It mainly contains tow components:
\begin{itemize}
  \item Trajectory Management: The trajectory of the laborer is represented by the AP-to-AP trajectory records (later introduced in Section \ref{sec_ap2ap}) in CHI. The AP-to-AP trajectories can be used to infer the relative location between APs, and in turn the APs of known location could help calibrating the AP-to-AP trajectories previously measured by the IMU sensors equipped in mobile phone. CHI collects the AP-to-AP trajectories during the time periods of the laborer's walking, evaluates the quality of each collected AP-to-AP trajectory, then generates and stores the compound trajectory of high quality by trajectory fusing. The details of the trajectory management is presented in Section \ref{tra_mgm}.
  \item Location-based Service: The service of locating the APs of interest, tracking the movement of the object, and drawing the the floor plan are provided in CHI. These services take the trajectory of the laborer and the location-based objectives set by the laborer as input, and offer the pathway suggestion and the solution of the objectives to the laborer. The details of the location-based service is presented in Section \ref{sec_lbs}.
\end{itemize}

Also, the laborer can set a list of location-based objectives, and re-order or change the objectives after then in CHI. A use case of CHI is presented in Section \ref{use_case}.

%In detail, CHI does not assume any priori knowledge (such as the location of the laborer, the location of APs, and the floor plan), and use IMU sensors to measure the direction and the distance of the laborer's movement. The relative location between APs can be inferred from the direction and the distance measured by IMU sensors, and in turn, the trajectory of the laborer can be calibrated by the location of an AP and the relative location between APs.

\section{Trajectory Management} \label{tra_mgm}
%Learned from previous studies [], the Inertial Measurement Unit (IMU) sensors equipped in mobile phone have poor performance in measurement. IMU-based trajectory tracking can suffer accumulate error over time due to the inaccuracy measurement of IMU. Therefore, we adopt AP-based calibration for trajectory management, and record AP-mark vector and AP-to-AP trajectories in CHI.
Learned from previous studies \cite{IMU_error}\cite{Escort}\cite{Pedestrian_tracking}, merely tracking the trajectory of the laborer by the IMU sensors equipped in mobile phone can suffer accumulate error over time, and thus cause poor performance on the measurement of the trajectory. Therefore, we introduce the notion of the AP-mark vector and AP-to-AP trajectory in trajectory management, and adopt AP-based calibration for the trajectory tracking in CHI.

\subsection{AP-Mark Vector}
\begin{figure} \centering
\subfigure[A AP-mark vector] { \label{fig2:a}
\raggedleft
\includegraphics[width=1.62in]{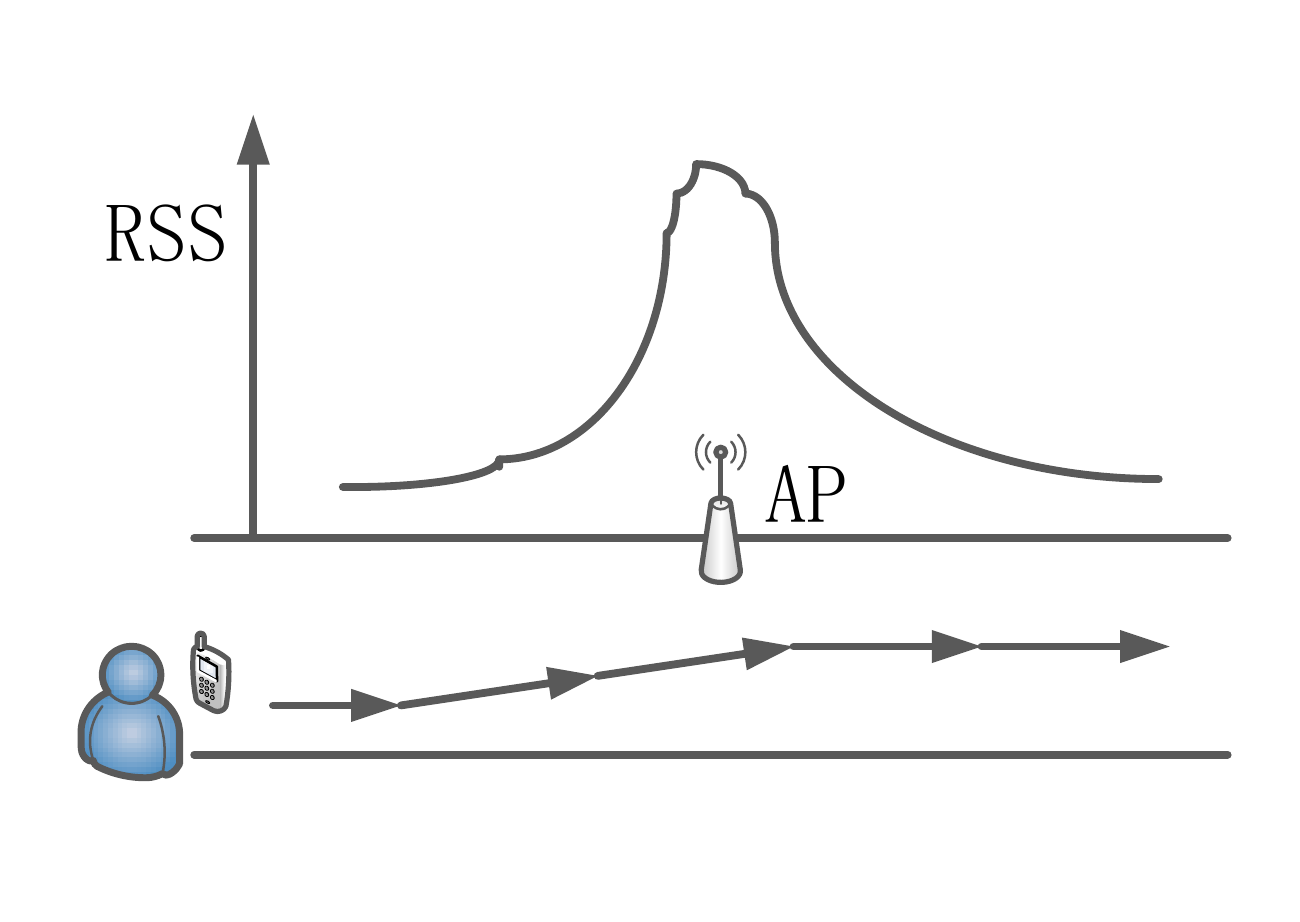}
}
\subfigure[Not a AP-mark vector] { \label{fig2:b}
%\centering
\raggedleft
\includegraphics[width=1.62in]{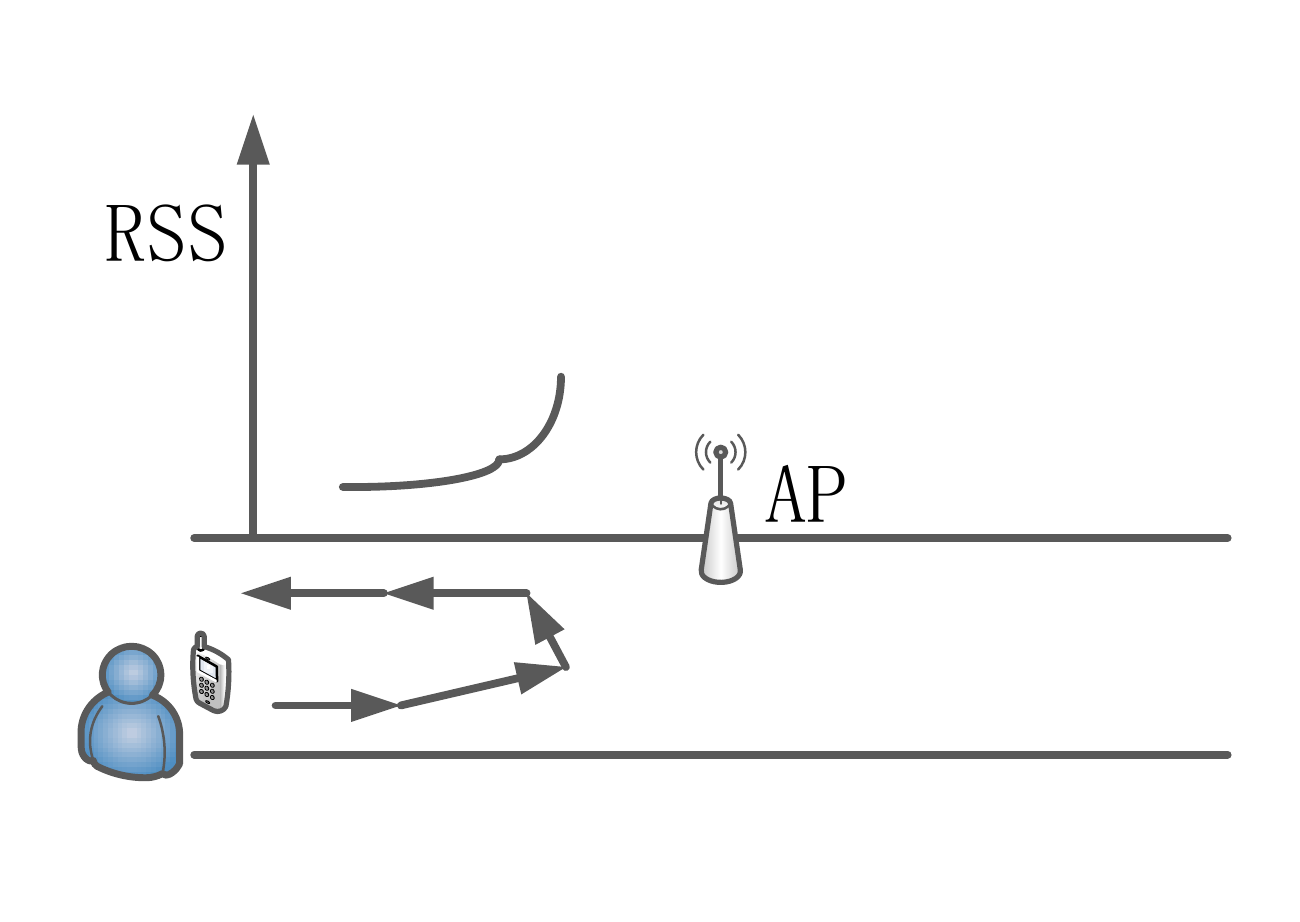}
}
\caption{The laborer walks to collect the RSS of signal from a AP}
\label{fig2}
\end{figure}

%The AP-mark Trajectory is based on the fact that when the laborer taking the phone straightly steps toward an AP, the RSS of the signal received by the phone from this AP gradually strengthens; when the laborer steps to the nearest location to the AP, the RSS received by the phone strengthens to the strongest; and then, when the laborer steps away from the AP, the RSS gradually weakens.
The AP-mark vector is based on the fact that when the laborer taking a phone walks approximately in a straight line and comes to the nearest location (named the AP-mark location) to an AP, the phone can receive the strongest signal sent by this AP (as shown in Figure \ref{fig2:a}). This fact has already been applied and shown its advantage on improvement of the localization accuracy in previous studies \cite{PI}\cite{Walkie-markie}. We can fit it to the CHI-based scenario.

In CHI-based scenario, the laborer does not always walk straightly ahead, e.g. taking a turn, and this walking manner may lead to a wrong determination of AP-mark location only by the signal strength such as shown in Figure \ref{fig2:b} that there exists a location has stronger signal compared to other locations in the laborer's trajectory. CHI uses the data of direction obtained by the gyroscope integrated in phone to distinguish the true or false AP-mark location. When the phone receives the strongest signal from an AP at a location, and (1) the change of its direction compared to the last locations is in a threshold value, e.g., within $\pm 20$ degrees in our implementation, CHI determines that this location is an AP-mark location, or (2) the change of its direction is greater than the threshold value, CHI determines that this location is not an AP-mark location.

When a location is determined to be an AP-mark location, CHI records the AP-mark vector that contains the data in locations with the change of their direction being within a threshold value (e.g. $\pm 20$ degrees) to the AP-mark location. In detail, the AP-mark vector includes the data records of timestamps, directions, the MAC address of the AP, the RSS of the signal from the AP, the MAC address of nearby APs, and the RSS of the signals from nearby APs at locations. Let the vector $\overrightarrow{v}$ represent the AP-mark vector. We have:
\begin{equation} \label{e1}
\overrightarrow{v}=<l_1,l_2,...,l_c>
\end{equation}
%\begin{equation} \label{e2}
%\begin{split}
%l_i=(Timestamp, \; MAC(AP), \; RSS(AP), \; \\ MAC(nearby \; APs) \; and \; RSS(nearby \; APs))
%\end{split}
%\end{equation}
Where we have $l_i=$($Timestamp$, $MAC(AP)$, $RSS(AP)$, $MAC(nearby$ $APs)$ and $RSS(nearby$ $APs))$, $i=1,2,...,c$.

\subsection{AP-to-AP Trajectory} \label{sec_ap2ap}
\begin{figure}
\centering
\includegraphics[width=2.2in]{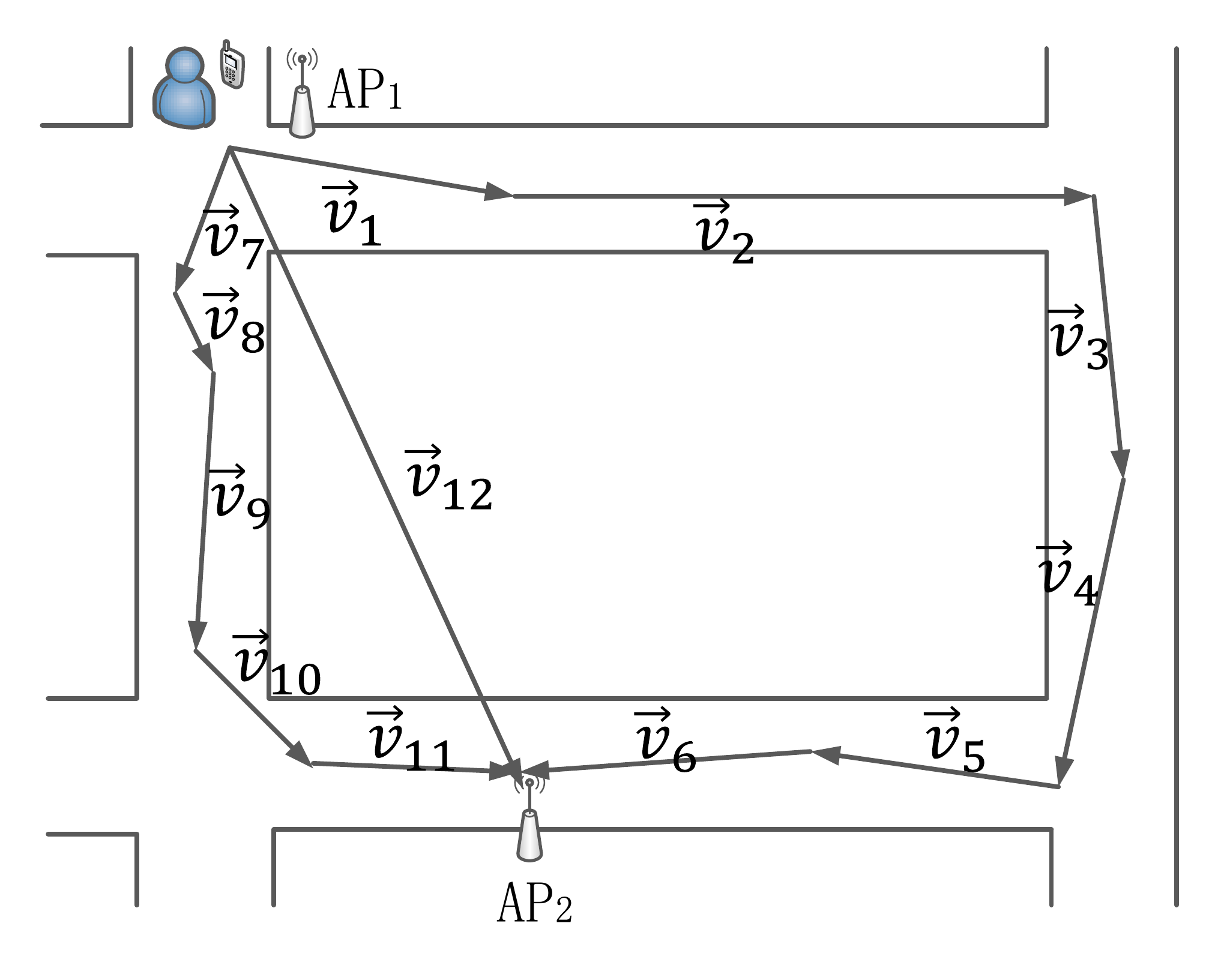}\\
\caption{\textrm{The AP-to-AP trajectories}} \label{fig3}
\end{figure}

AP-to-AP trajectory, which can be used to estimate the relative location between APs, is the laborer's walking trajectory between two AP-mark vectors that does not go through any other AP(, or to say, does not contain the AP-mark vector of any other AP) except these two APs. AP-to-AP trajectory can be represented by a group of vectors, and then the displacement between APs can be estimated by summing up this group of vectors. For example, in figure \ref{fig3}, two group of vectors, $<\overrightarrow{v}_1,\overrightarrow{v}_2,\overrightarrow{v}_3,\overrightarrow{v}_4,\overrightarrow{v}_5,\overrightarrow{v}_6>$ and $<\overrightarrow{v}_7,\overrightarrow{v}_8,\overrightarrow{v}_9,\overrightarrow{v}_{10},\overrightarrow{v}_{11}>$, represent two trajectories between $AP_1$ and $AP_2$ respectively, and the displacement $\overrightarrow{v}_{12}$ between $AP_1$ and $AP_2$ has $\overrightarrow{v}_{12}=\sum_{i=1}^{6} \overrightarrow{v}_i$ or $\overrightarrow{v}_{12}=\sum_{i=7}^{11} \overrightarrow{v}_i$.

In implementation, the locations within a threshold direction change (e.g. within $\pm 20$ degrees) from a start point location on laborer's trajectory are considered to be to the locations that compose of a vector. The direction of the vector can be calculated by invoking the Android API \texttt{getRotationMatrixFromVector()}, which provides direction estimation based on Kalman Filter \cite{Kalman}, and has been adopted by many apps. CHI uses the default parameters of \texttt{getRotationMatrixFromVector()} to estimate the direction of the vector. The length of the vector can be calculated by directly applying the data from the accelerometer integrated in phone. CHI uses the NASC algorithm \cite{Zee} reported in \cite{Step_comparison} that has better performance than other algorithms to count the steps of the laborer, and estimate the length of the vector by training the linear relationship between step frequency and stride length \cite{Step_length}.

\subsection{Trajectory Fusing} \label{sec_tf}
With the laborer's walking, the collected AP-to-AP trajectories may have overlapping vectors between each other, e.g., two AP-to-AP trajectories containing an overlapping pathway at the same passage. In this case, a part of vectors out of these overlapping vectors are needed to be selected and fused to construct a compound high-quality vector. In implementation, CHI applies the MCD estimator \cite{MCD} to fuse the overlapping vectors. Suppose that the vector $\overrightarrow{v}_1,\overrightarrow{v}_2,...,\overrightarrow{v}_n$ are the overlapping vectors in $d$-dimensional space. MCD estimator in CHI-based scenario solves the following optimization problem:
\begin{equation} \label{e2}
\begin{cases}
\displaystyle \text {min: } det(cov(\overrightarrow{v})) \\ \\
\displaystyle \text {s.t. \;} \overrightarrow{v} = \frac{\sum_{i=1}^{n} w_i \overrightarrow{v}_i}{\sum_{i=1}^{n} w_i}, \\
\displaystyle \text {\;\;\;\;\;} cov(\overrightarrow{v}) = \frac{\sum_{i=1}^{n} w_i(\overrightarrow{v}_i-\overrightarrow{v}) (\overrightarrow{v}_i-\overrightarrow{v})^{'}}{\sum_{i=1}^{n} w_i - 1}, \\
\displaystyle \text {\;\;\;\;\;} w_i=0 \;\text{or}\; 1, \; \sum_{i=1}^{n} w_i = h, \; \text{and} \; h=\lfloor\frac{n+d+1}{2}\rfloor
\end{cases}
\end{equation}
Where a total of $h$ vectors are selected to minimize the determinant of $cov(\overrightarrow{v})$. CHI applies Fast-MCD algorithm \cite{Fast_MCD} in MCD estimator to calculate this optimization problem, and the vector $\overrightarrow{v}$ in (\ref{e2}) is the compound vector after fusing.

In addition, since CHI is implemented in mobile phone, we manage to save the computation and storage cost. When CHI applies the trajectory fusing process and $n>10$ overlapping vectors exists in a trajectory, CHI keeps the $h=\lfloor\frac{n+d+1}{2}\rfloor$ vectors selected by the MCD estimator and discards other unselected vectors.

\begin{figure*}[!htb] \centering
\subfigure[Ground truth] { \label{fig4:a}
\raggedleft
\includegraphics[width=1.58in]{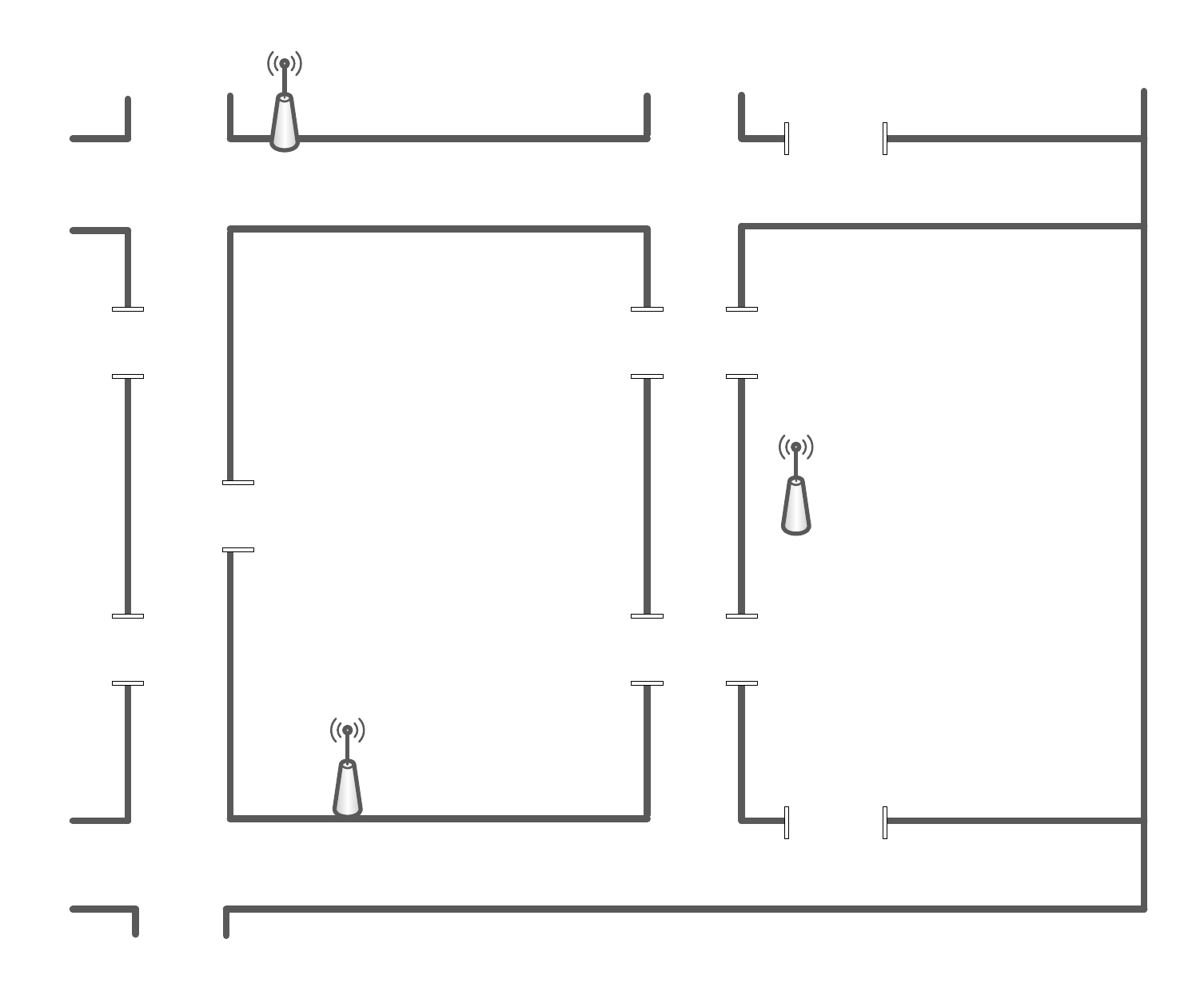}
}
\hspace{0.05in}
\subfigure[Assigning Points to be covered] { \label{fig4:b}
\raggedleft
\includegraphics[width=1.58in]{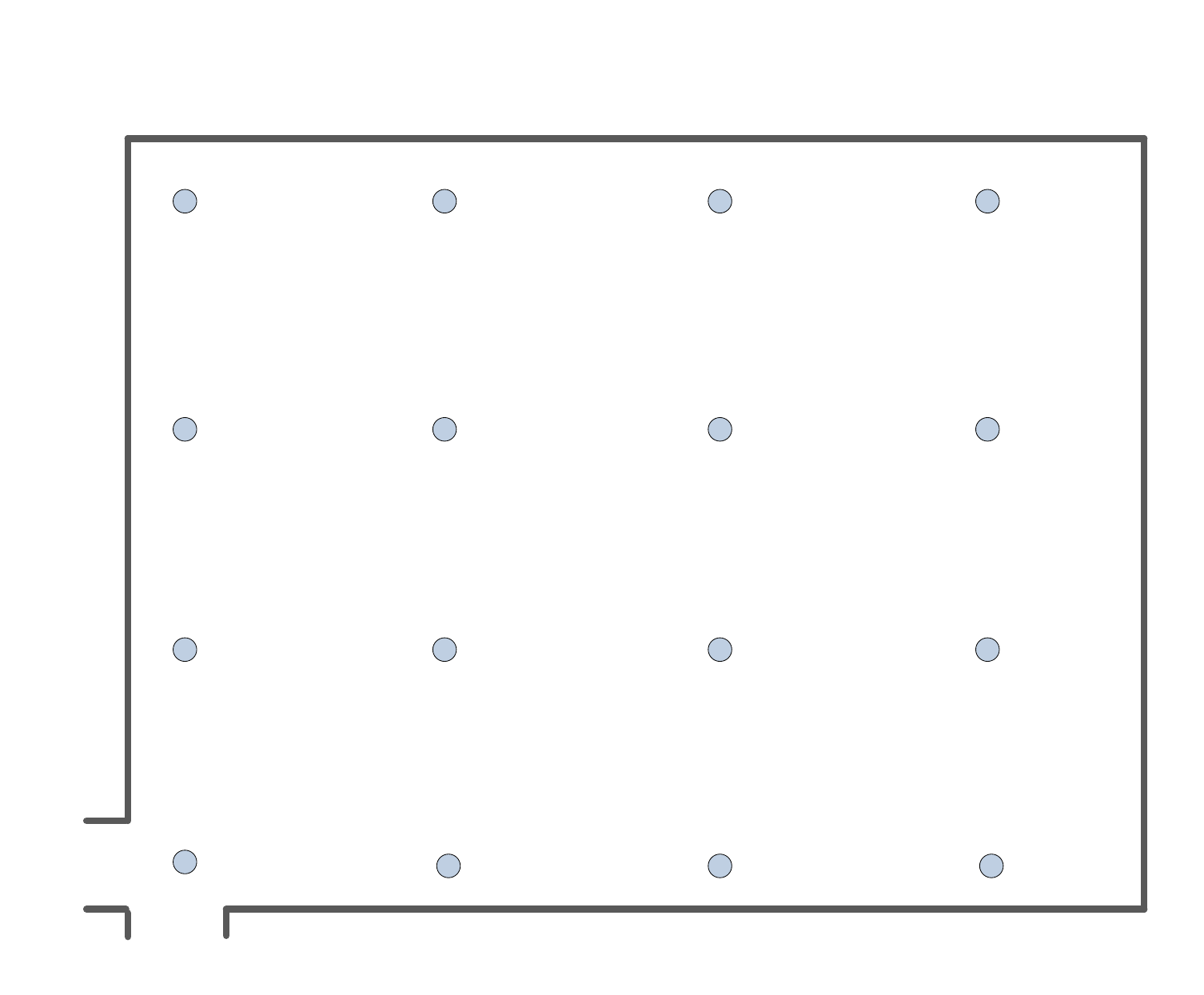}
}
\hspace{0.05in}
\subfigure[Suggesting the pathway] { \label{fig4:c}
%\centering
\raggedleft
\includegraphics[width=1.58in]{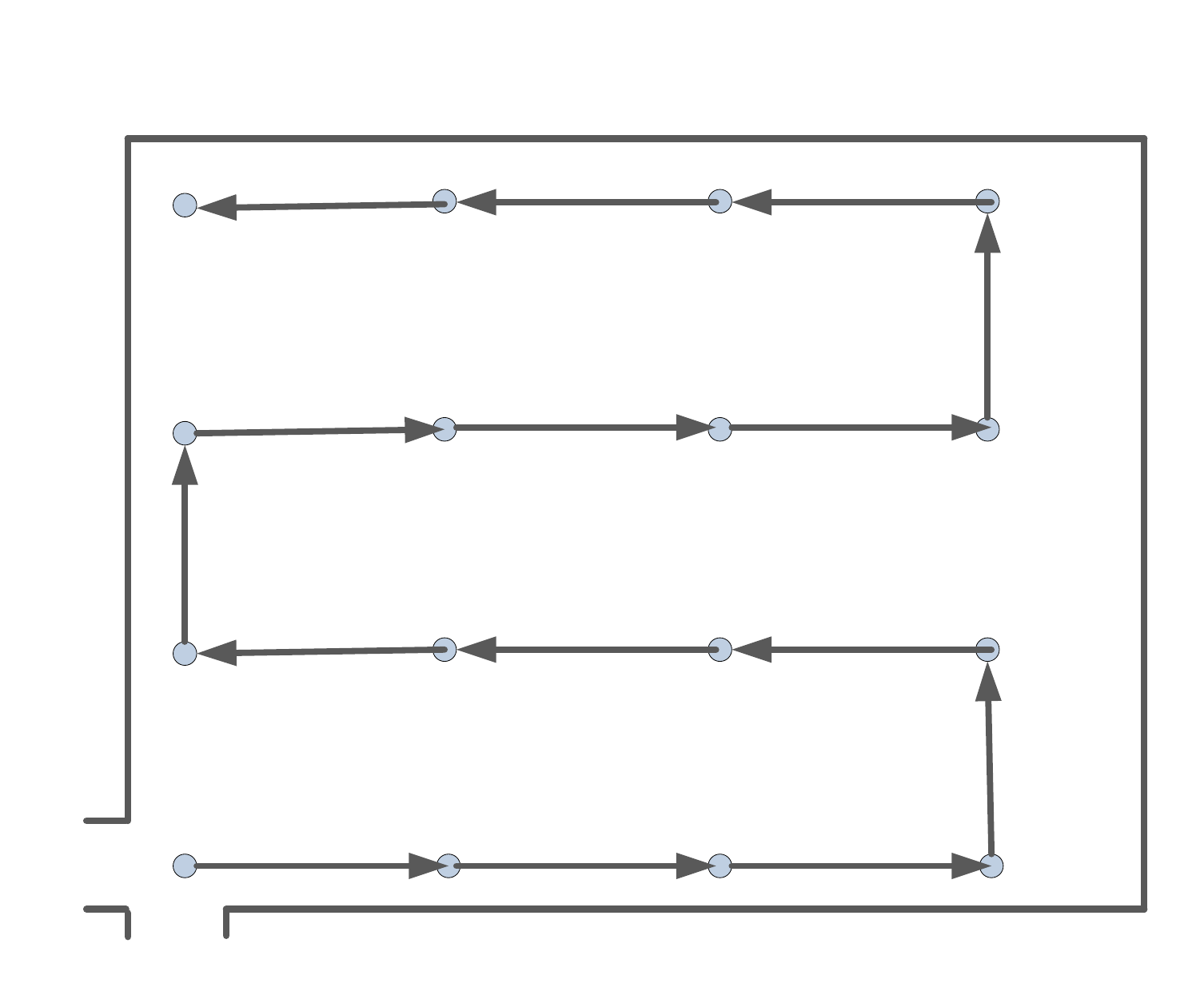}
}
\hspace{0.05in}
\subfigure[The pathway being split into components] { \label{fig4:d}
\raggedleft
\includegraphics[width=1.58in]{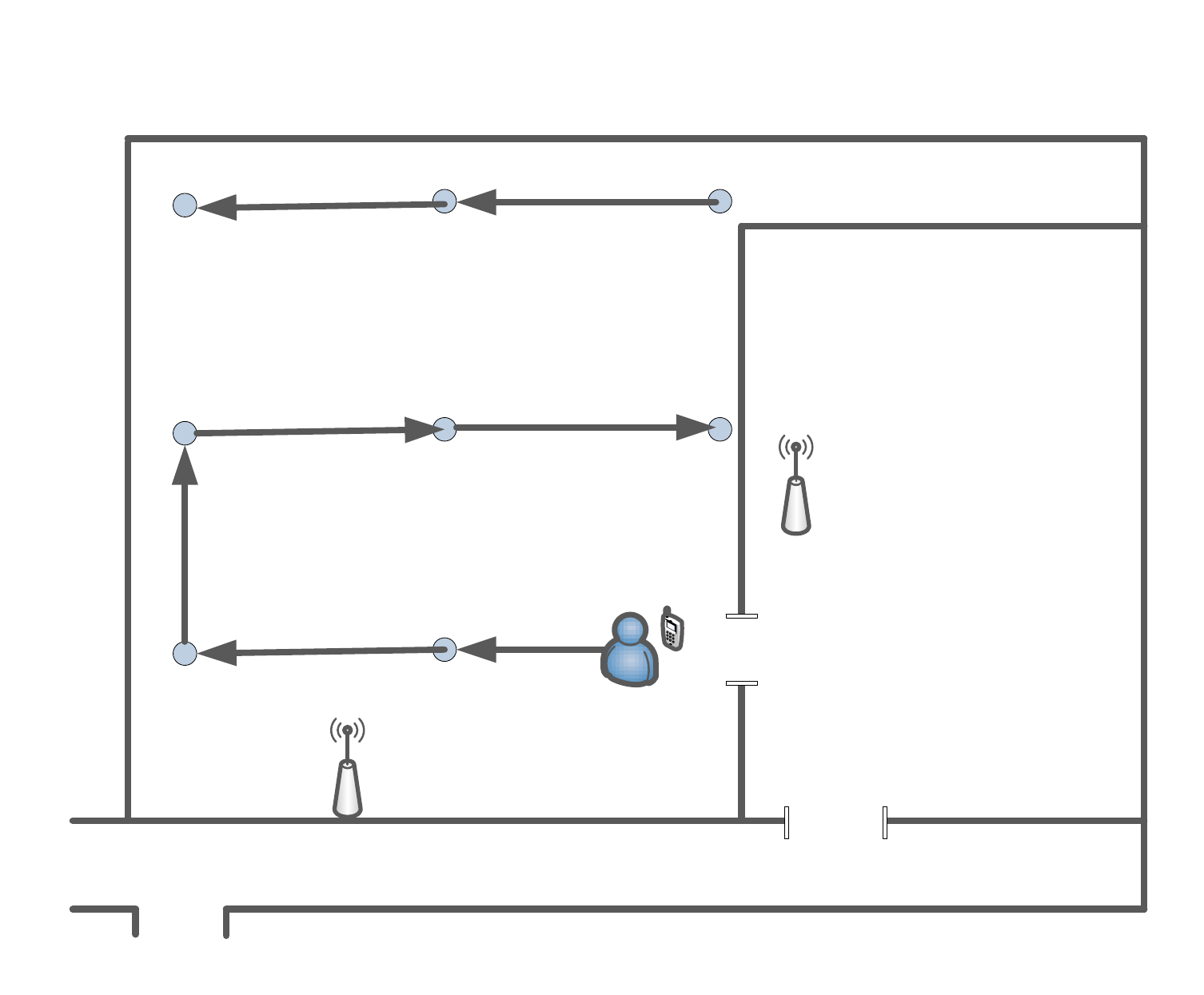}
}
\subfigure[Several AP-to-AP trajectories may existing between the same pair of APs] { \label{fig4:e}
\raggedleft
\includegraphics[width=1.60in]{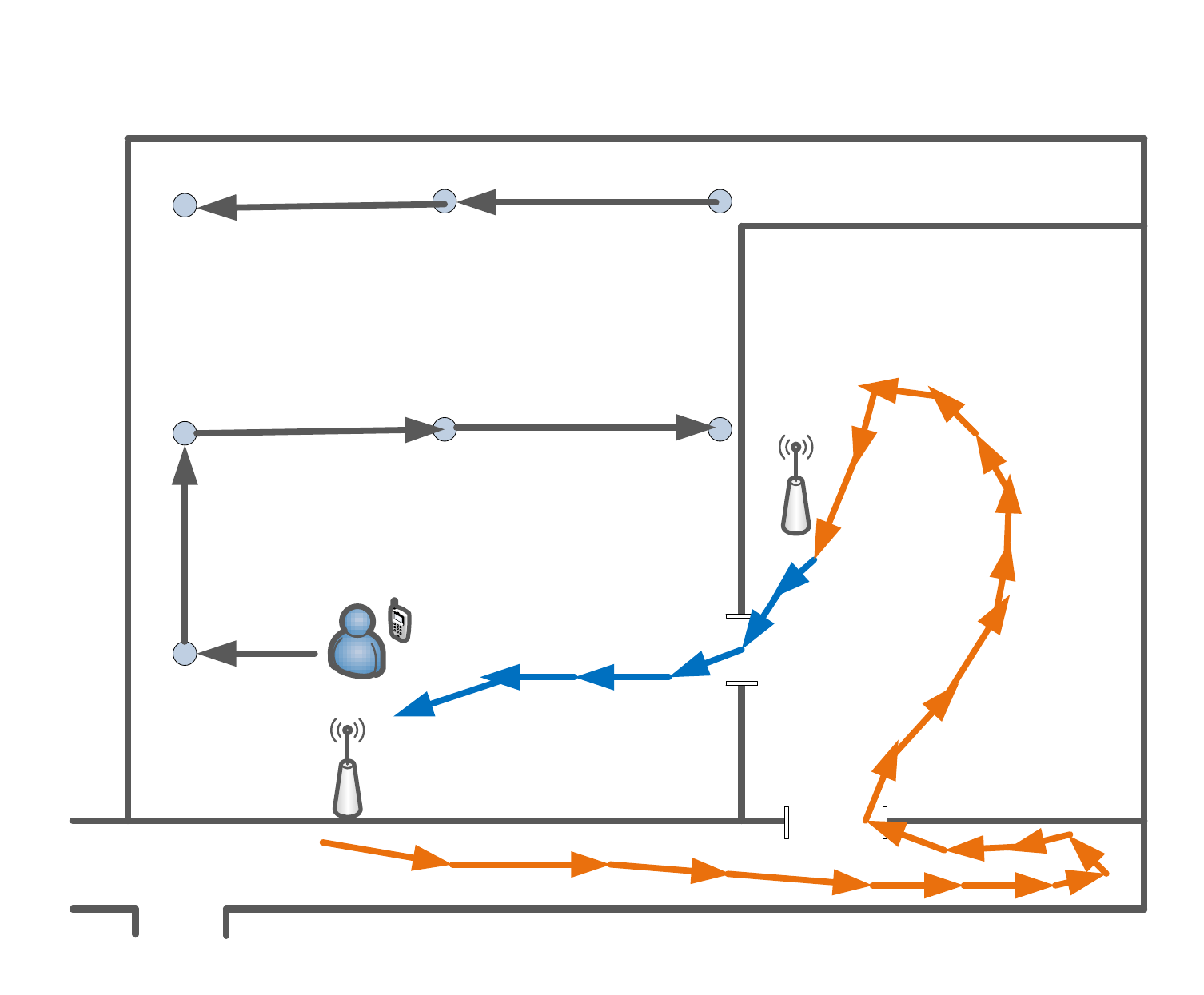}
}
\hspace{0.05in}
\subfigure[Dividing the area around an AP into $8$ sectors] { \label{fig4:f}
\raggedleft
\includegraphics[width=1.58in]{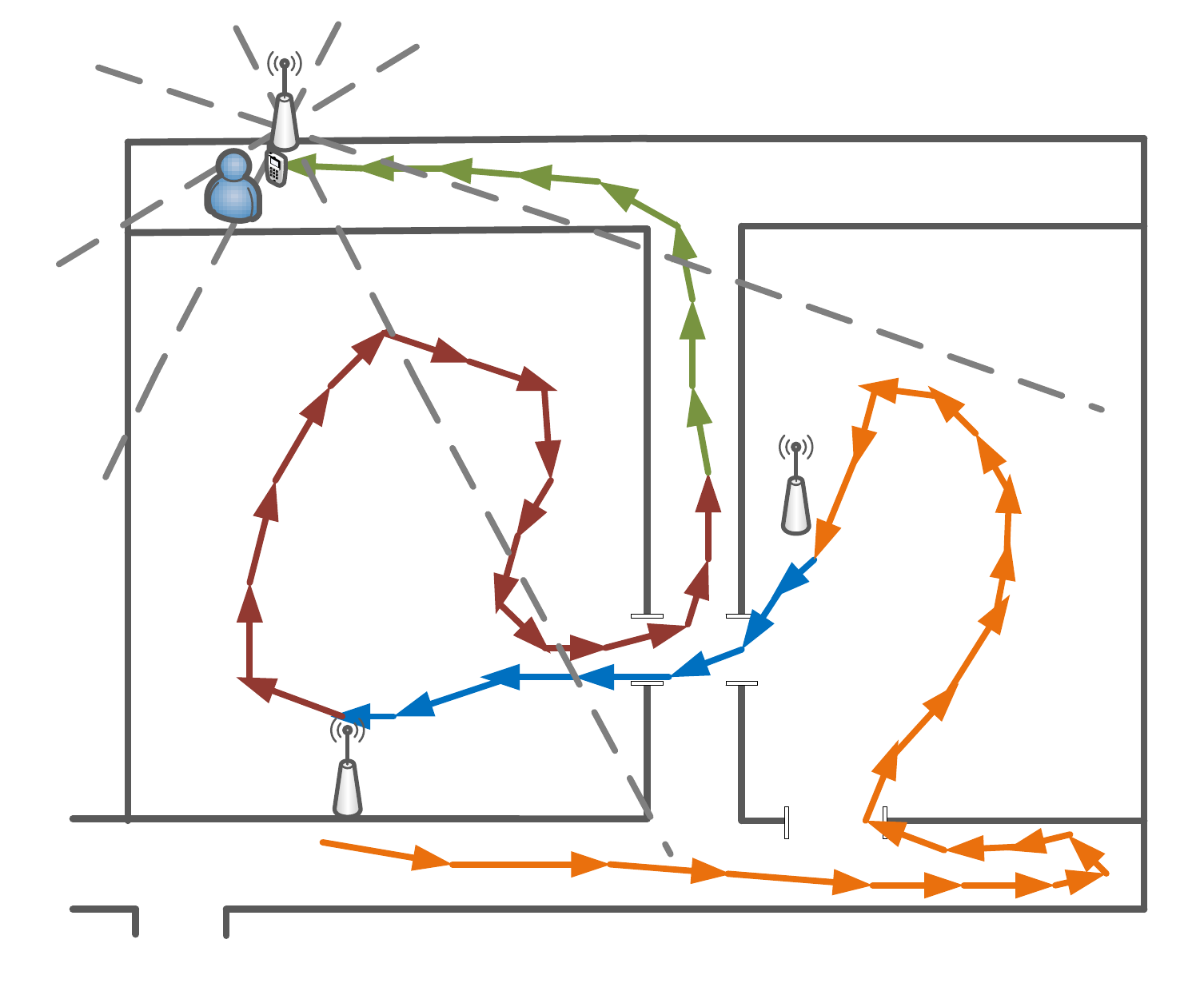}
}
\hspace{0.05in}
\subfigure[Two closed paths] { \label{fig4:g}
\raggedleft
\includegraphics[width=1.58in]{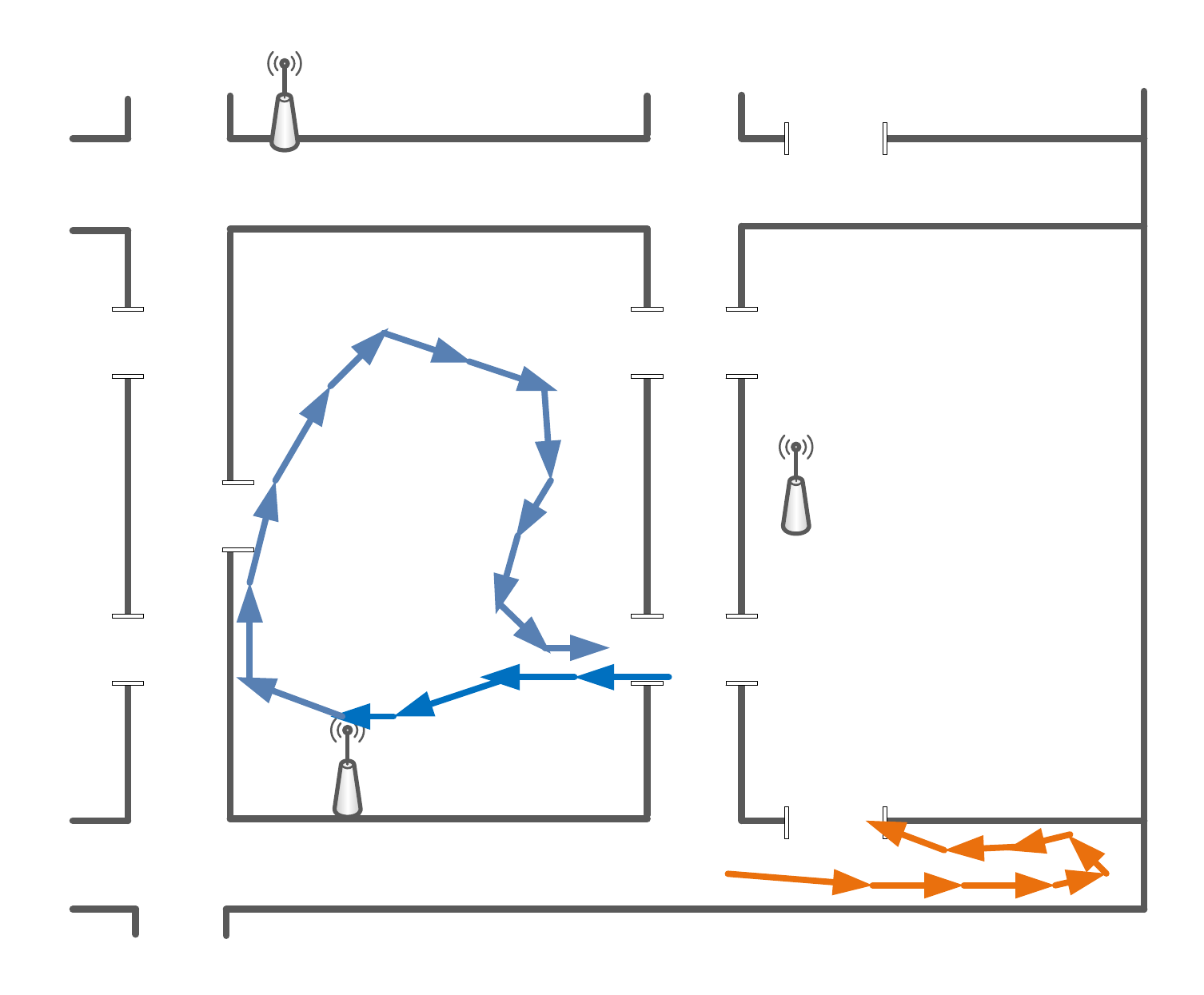}
}
\hspace{0.05in}
\subfigure[Two paths with a turn] { \label{fig4:h}
\raggedleft
\includegraphics[width=1.58in]{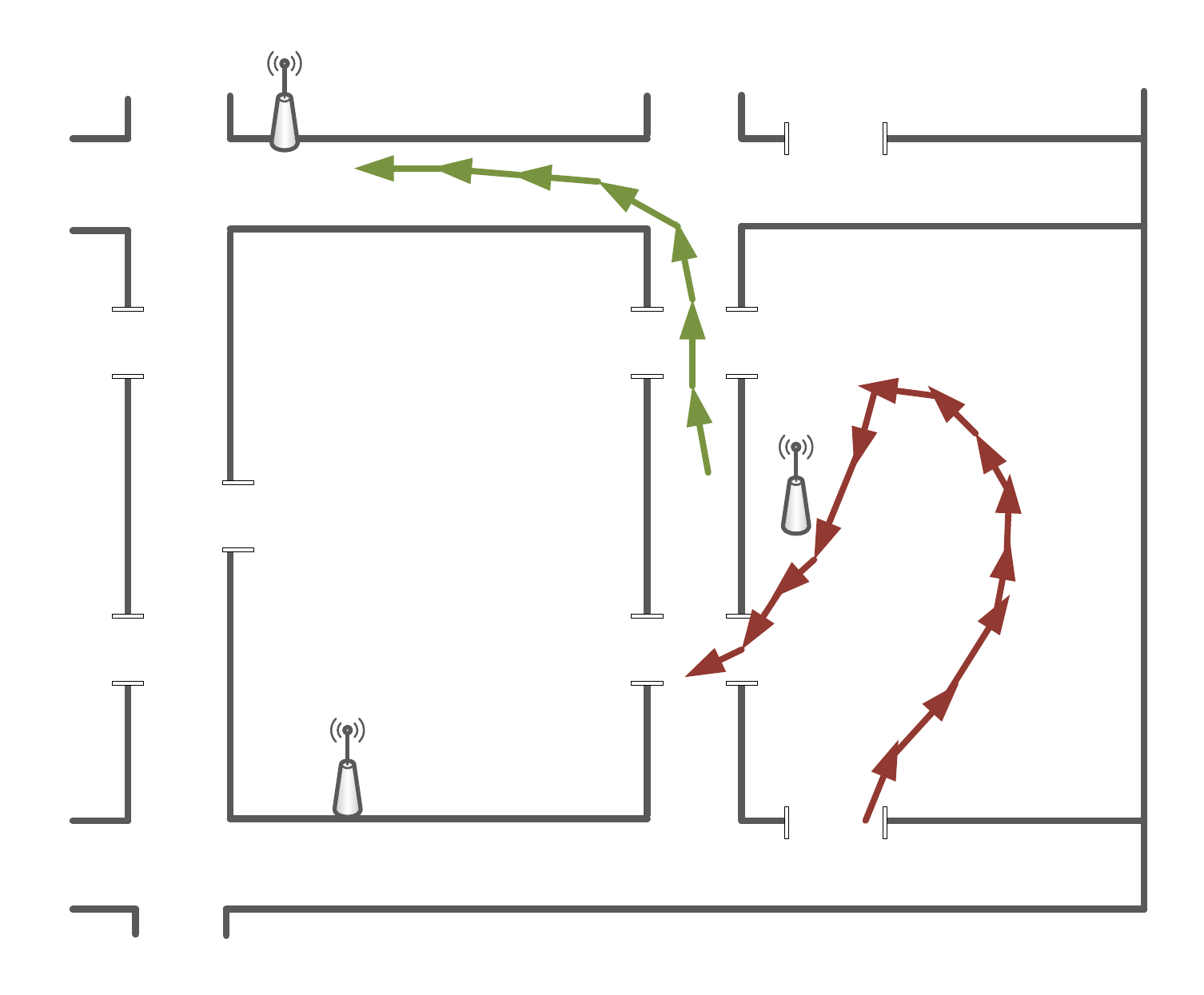}
}
\caption{Example of location-based services}
\label{fig4}
\end{figure*}

\section{Location-based Service} \label{sec_lbs}
Surveying previous studies, e.g., \cite{Model_AP}\cite{FootSLAM}\cite{Unloc}, there are mainly three basic location-based services: (1) locating the APs of interest, (2) tracking the movement of the object, and (3) drawing the floor plan. CHI implements these three location-based services in cyber-human interaction fashion that the laborer can manually set a list of objectives on the service, and the location-based service suggests the pathway for the laborer while providing the solution for the list of the objectives.

\subsection{Locating The APs of Interest} \label{sec_ap}
The laborer can indicate the APs of interest by directly selecting APs in the list, designating an area, or setting to explore all possible existing APs in the area of the floor in CHI, and CHI implements two sub-services for the APs of interest: sub-service 1 --- providing pathway suggestion for finding all APs in the indicated area and positioning the relative locations between APs, and sub-service 2 --- providing path suggestion for improving the quality of trajectories between APs and re-positioning the relative locations between APs.

\subsubsection{Sub-service 1} \label{sec_subs1}
CHI firstly assign a set of uniformly distributed points on the area (as shown in Figure \ref{fig4:b}), and set the distance between two adjacent (left-right and up-down) points to be the coverage radius of WiFi signal. The target of pathway suggestion in this sub-service is to find the shortest Hamilton path so that the laborer can walk following this path to find all possible APs in the area with least effort. When there is no obstacle in the area, CHI can find the shortest Hamilton path like Figure \ref{fig4:c} for pathway suggestion. As the laborer walks, CHI removes the points that have been visited(, or to say, points that are within the distance of the WiFi signal's coverage radius to the trajectory of the laborer) in the shortest Hamilton pathway of suggestion. Here, two cases happening will split the pathway into unconnected components. The first case is that the laborer does not walk following the pathway suggestion, and the second case is that some obstacles have been found by the floor plan process (introduced in Section \ref{sec_floorplan}), and CHI removes parts of the pathway that go through the obstacles. When the pathway is split into components, CHI does not try to re-establishes the connection between components as we find that it would puzzle the judgement of the laborer for pathway choice in implementation.

When all points in the pathway suggestion have been visited or the laborer terminates the process, CHI starts to position the relative locations between APs by using the collected AP-mark vectors and AP-to-AP trajectories. Two APs may have several AP-to-AP trajectories in between (e.g. the orange trajectory and the blue trajectory in Figure \ref{fig4:e}), and CHI uses the start or end AP-mark vector to distinguish these AP-to-AP trajectories and simply choose the trajectory with the least length for positioning since the trajectory with less length has less accumulated error. Then, the relative locations between APs can be calculated by existing positioning algorithms. In CHI, we implement the Arturia algorithm proposed in Walkie-Markie \cite{Walkie-markie}, whose experiment shows that Arturia has better performance compared to other algorithm, for localization.

\subsubsection{Sub-service 2} \label{sec_subs2}
In this sub-service, CHI gives two types of pathway suggestion. First, CHI divides the area around each AP into eight sectors with equal angle (as shown in Figure \ref{fig4:f}). If there exists a neighbor AP in one of the sectors of an AP, and the AP-to-AP trajectories between these two APs\footnote{Note that not merely one AP-to-AP trajectory exists between two APs. With different AP-mark vectors as the start point from an AP, the AP could establish two or more AP-to-AP trajectories with another AP. In CHI, we recommend the laborer to collect two orthogonal AP-mark vectors for an AP.} have not yet been collected, CHI would try to find a pathway between these two APs that does not go through any other AP(, or to say, does not contain the AP-mark vector of any other AP) and is not blocked by any obstacle. If such a pathway could be found, CHI recommends it to the laborer. Second, for these AP-to-AP trajectories used in the Arturia positioning algorithm, CHI would suggest the laborer to retrace them again for trajectory fusing in order to improve the positioning accuracy. As the laborer retraces the suggested pathway of AP-to-AP trajectories, the trajectory management component in CHI would repeatedly executes the fusing process on each AP-to-AP trajectory until the displacement vector $\overrightarrow{v}_{(i+1)}$ between two APs has $|\overrightarrow{v}_{(i+1)}-\overrightarrow{v}_{(i)}|<\theta$ (e.g. $\theta<1$m) in the $i+1$th fusing or the laborer can terminate this process. After the pathway suggestion process has been finished or terminated, CHI re-applies the Arturia algorithm to calculate out the relative locations between APs.

\subsection{Tracking The Movement of The Object}
%初始时，the laborer的路径用通过手机IMU获取到的数据来估计，当通过一段时间的行走获取到了AP-marked trajectroy和AP-to-AP trajectory时，则进行校正。In details,
%1) 用户首先选取要track的路径的时间范围和区域范围
%2）如果经过的AP的相对位置并未计算出，则采用手机IMU获取到的数据来估计
%3）如果经过的AP的相对位置已经计算出，则用AP-marked trajectory和AP-to-AP trajectory 来校准。
%收集要track的路径所经过的为计算出相对位置的AP，并建议the laborer定位这些AP；如果the laborer接受了这一建议，那么则圈定一个以这些AP做为顶点多边形区域，用上一个section所描述的方法为用户提供建议路径，并计算这一个区域内所有AP（包括已经计算出相对位置的AP）的相对位置
%另外，如前面的section所述，随着trajectory变多，一些trajectory可能因为fusing而被去掉，因此被删掉trajectory的时间戳会被打到fusing出来的trajectory以替代原来的trajectory for query.
Initially, CHI tracks the movement of the laborer through the data obtained by IMU sensors as introduced in Section \ref{sec_ap2ap}. Then, after a time period of the laborer's walking and collecting, CHI can calibrate the movement trajectory of the laborer by the established relative locations between APs. In details:
\begin{enumerate}[]
  \item The laborer specifies a range of time and an area, and CHI starts to track the trajectory of the laborer with the time and area specification.
  \item If the trajectory went through an AP, and the relative location of this AP is unknown, CHI uses the data obtained by IMU sensors to estimate the trajectory of the laborer.
  \item If the trajectory went through an AP, and the relative location of this AP is known, CHI uses the collected AP-mark vectors and AP-to-AP trajectories to calibrate the estimation of the laborer's trajectory.
\end{enumerate}

For improving the tracking accuracy, CHI would suggest the laborer to position the relative locations of the APs that the tracked trajectory went through. If the laborer accepts this suggestion, CHI constructs a polygon area with all these APs located in, and provides the same pathway suggestion as the suggestion in Section \ref{sec_ap}. When all points in the pathway suggestion have been visited or the laborer terminates the process, CHI can calculate out the relative locations between APs to help calibrating the movement trajectory of the laborer. In addition, as more trajectories are collected with the laborer's movement, some of trajectories would be discarded by the trajectory fusing process (addressed in Section \ref{sec_tf}). In this situation, CHI simply tags the timestamp of discarded trajectories to the compound trajectory for query.

\subsection{Drawing the Floor Plan} \label{sec_floorplan}
To draw the floor plan, the laborer are firstly required to estimate and input the length and width of the floor to CHI. Then, CHI applies the similar method used in Section \ref{sec_subs1} that assigns a set of uniformly distributed points on the area of the floor and constructs the shortest Hamilton path for pathway suggestion. Unlike the method in Section \ref{sec_subs1}, the distance between two adjacent points should be set to a smaller value (e.g. $2$m) since the floor plan requires more intensive detection of objects than detection of APs.

CHI considers three basic component categories of Passage, Entrance and Room in drawing the floor plan. Initially, CHI sets all components along the trajectory of the laborer as Passages in default, and introduces the following judgement statements for correction:
\begin{enumerate}[]
  \item If there is a closed path in the trajectory, and this closed path can be divided into two overlapping sub-paths (or two sub-paths are quite close), CHI determines that there is a dead-end in the Passage and add a block to the dead-end (e.g. the orange path in Figure \ref{fig4:g}).
  \item If there is a closed path in the trajectory, and this closed path cannot be divided into two overlapping sub-paths, CHI determines that the area covering the closed path is a Room and add an Entrance in the closed point (e.g. the blue path in Figure \ref{fig4:g}). This statement applies to the case that a room has only one Entrance or the laborer enters and leaves the room through a same Entrance.
  \item If there is a turn, whose length exceeds a threshold value (e.g. $5$m), in the trajectory, and the turn cannot be approximately represented by a broken line(, or to say, two vectors), CHI determines that the area covering the turn is a Room and add two Entrances in both ends of the turn. (For example, CHI determines the area covering the red path is a room while the area covering the green path is not a room in Figure \ref{fig4:h}.) This statement applies to the case that the laborer enters and leaves the room through two different Entrances.
\end{enumerate}

However, due to the measurement error and the randomness of the laborer's walking, misjudgments are inevitable in drawing the floor plan. The laborer can manually correct and draw the component in CHI. Once the laborer locks a component, CHI holds the state of the component and no longer modify it.

\begin{figure*}
\centering
\includegraphics[width=5.2in]{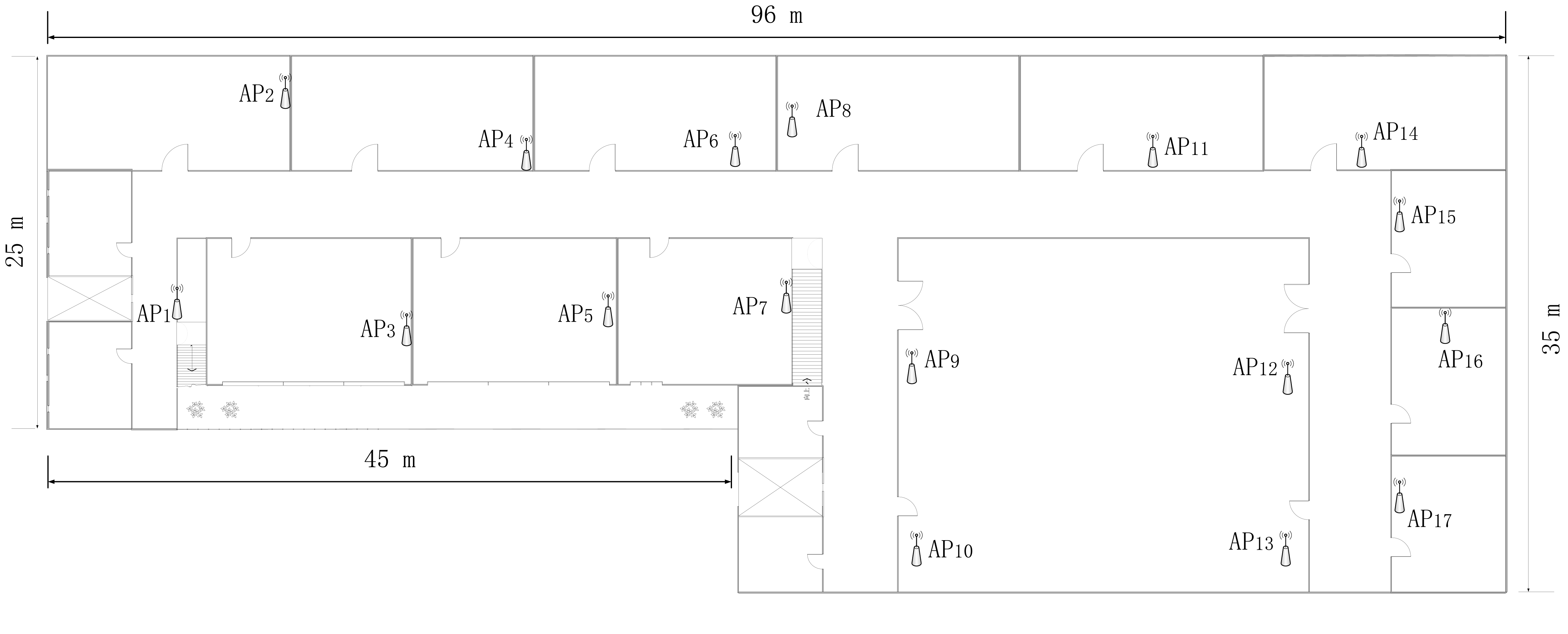}\\
\caption{\textrm{Floor plan}} \label{fig5}
\end{figure*}

\begin{figure*}[!htb] \centering
\subfigure[Suggesting the pathway for locating the APs of interest] { \label{fig6:a}
\centering
\includegraphics[width=1.5in,height=2.8in]{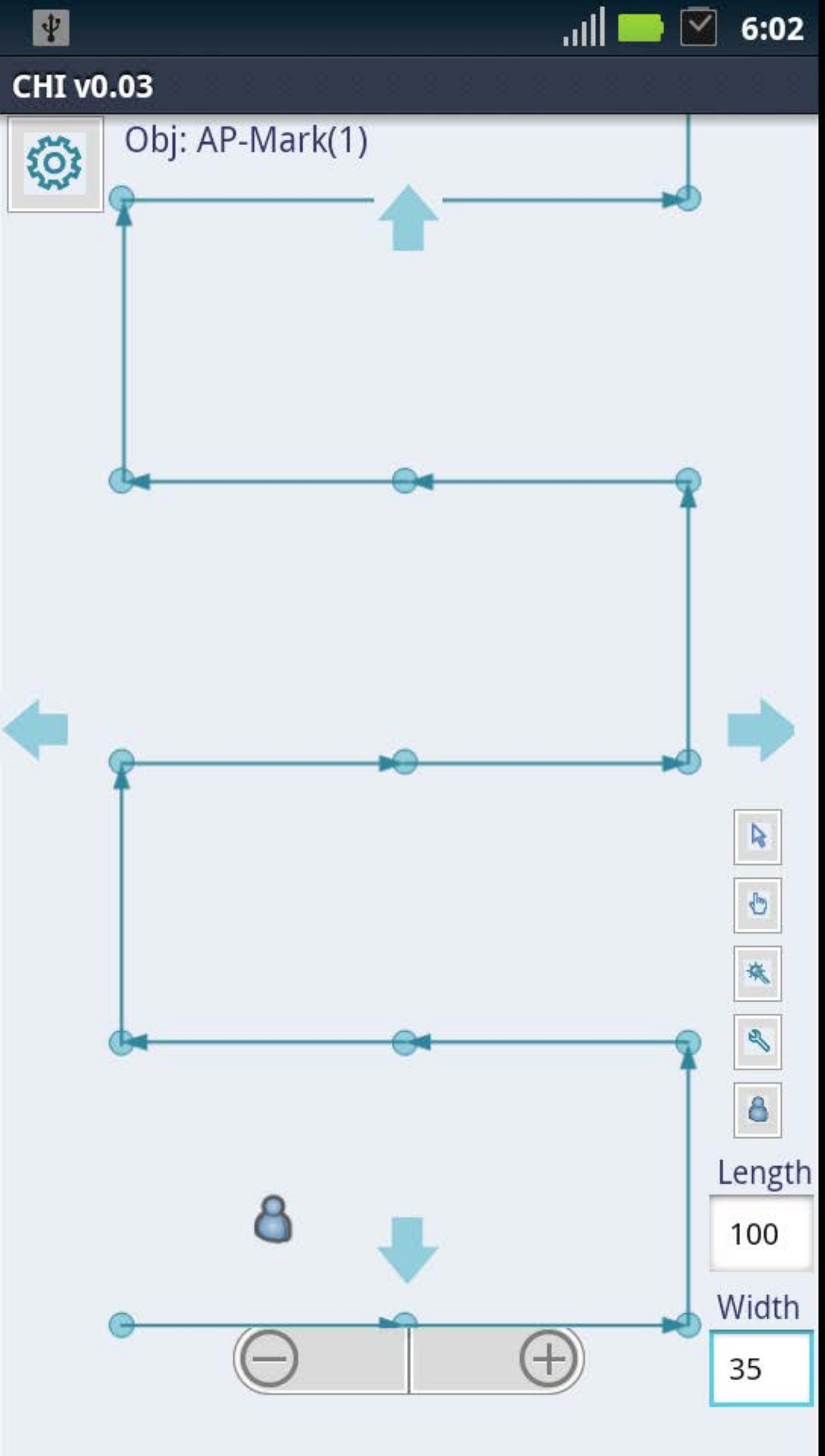}
}
\hspace{0.05in}
\subfigure[Three rooms having been found] { \label{fig6:b}
\centering
\includegraphics[width=1.5in,height=2.8in]{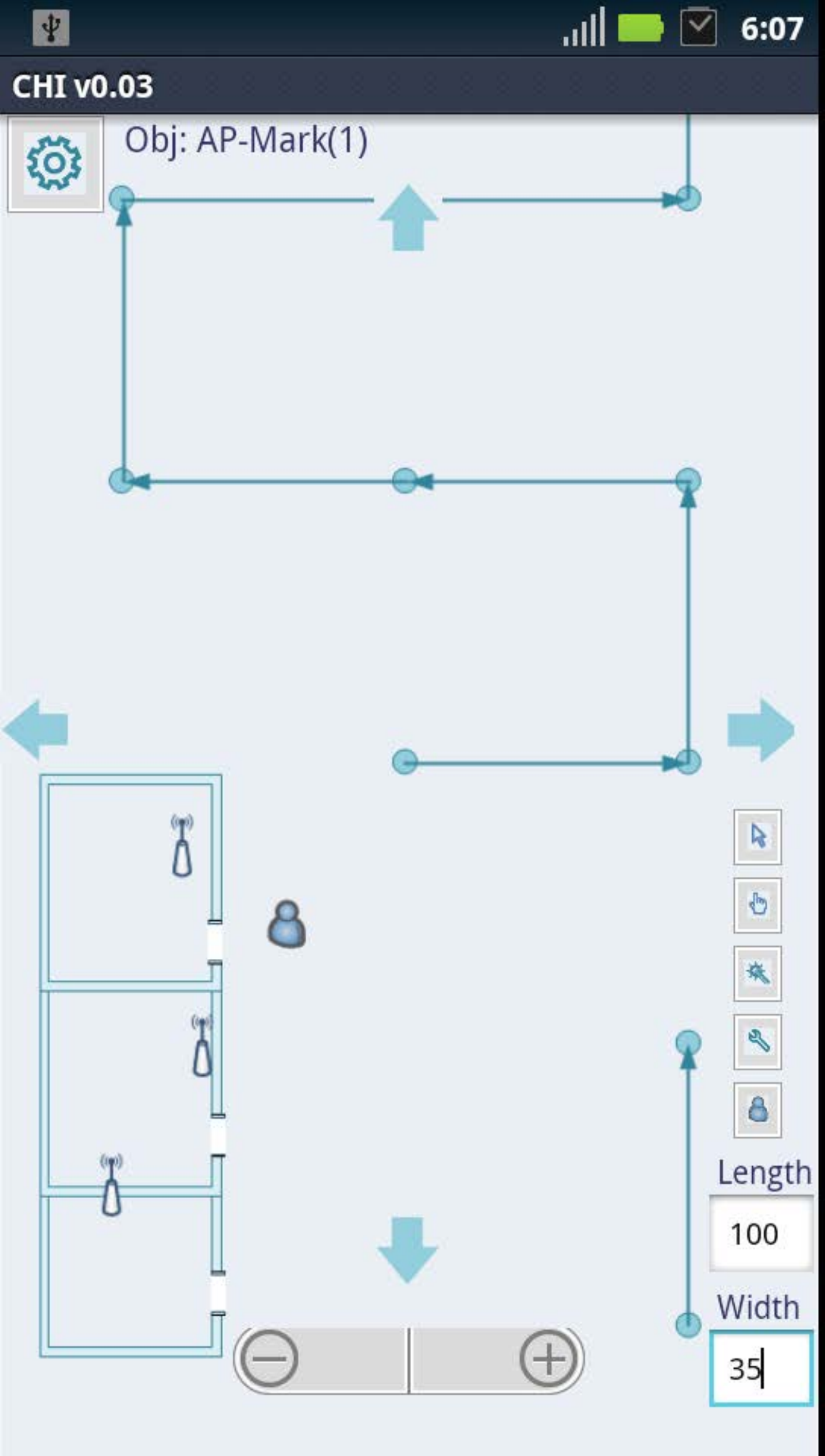}
}
\hspace{0.05in}
\subfigure[Changing the primary objective to the floor plan] { \label{fig6:c}
%\raggedleft
\centering
\includegraphics[width=1.5in,height=2.8in]{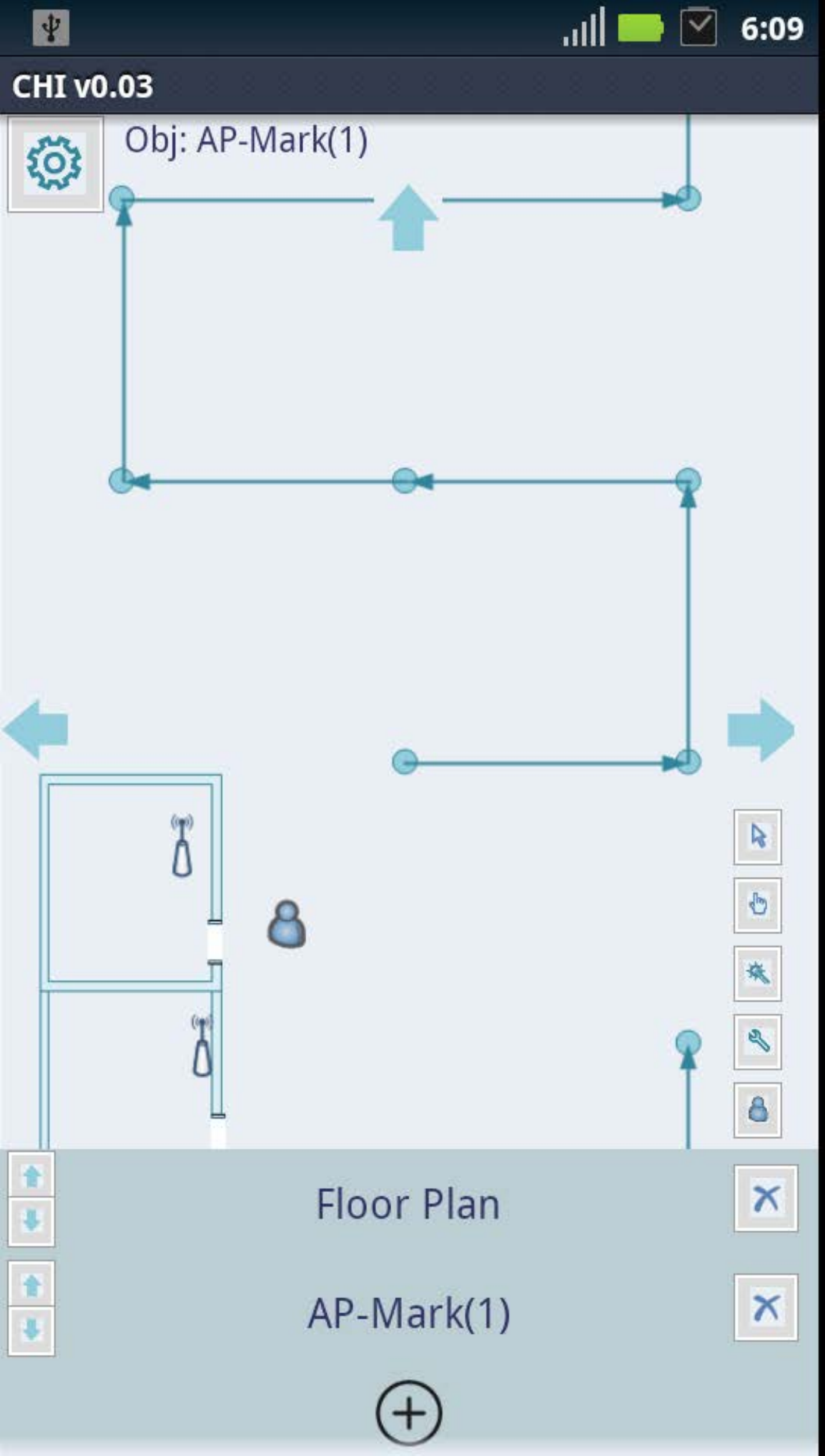}
}
\hspace{0.05in}
\subfigure[Suggesting the pathway for drawing the floor plan] { \label{fig6:d}
\centering
\includegraphics[width=1.5in,height=2.8in]{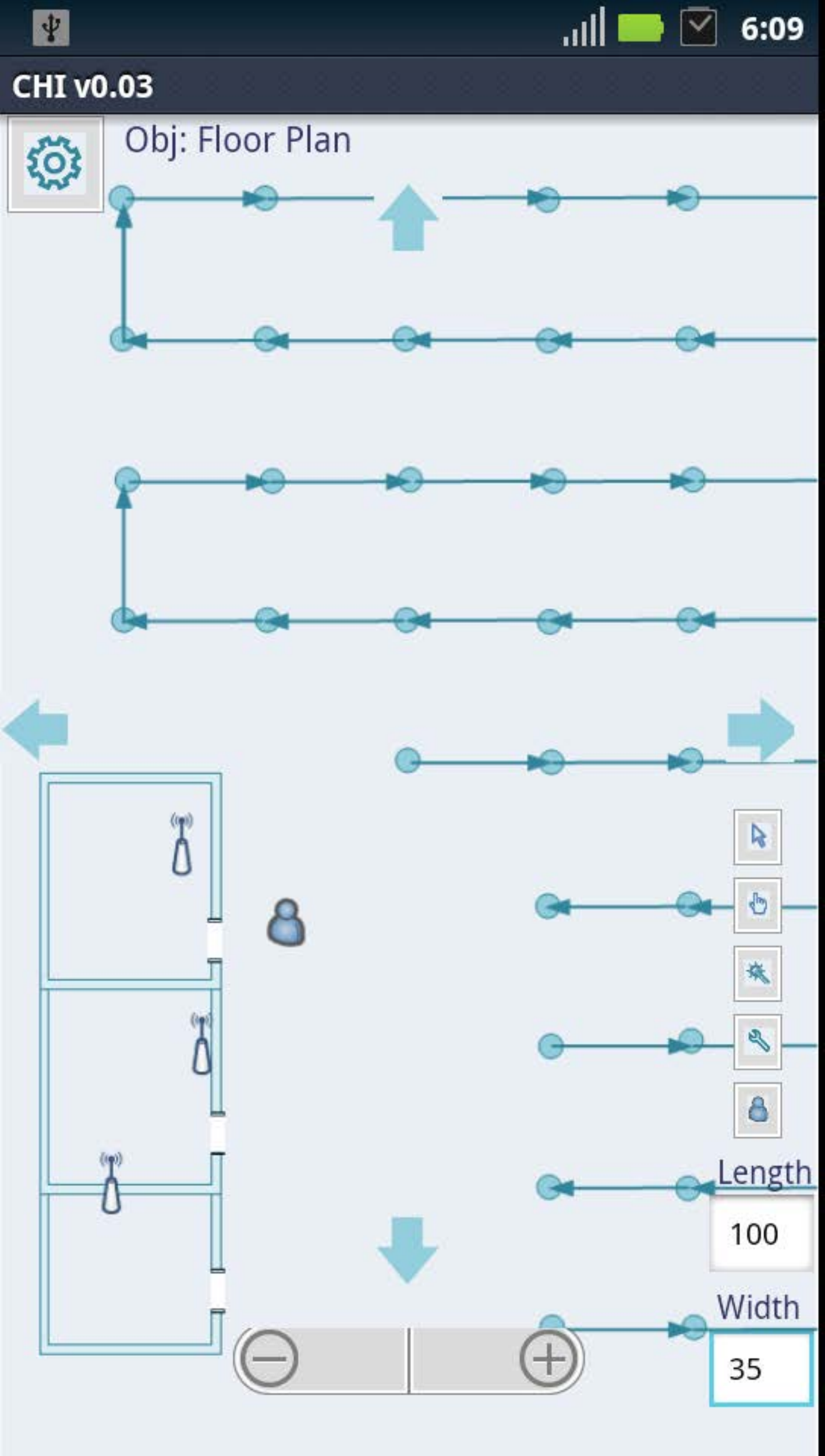}
}
\subfigure[The raw result of the floor plan] { \label{fig6:e}
\centering
\includegraphics[width=1.5in,height=2.8in]{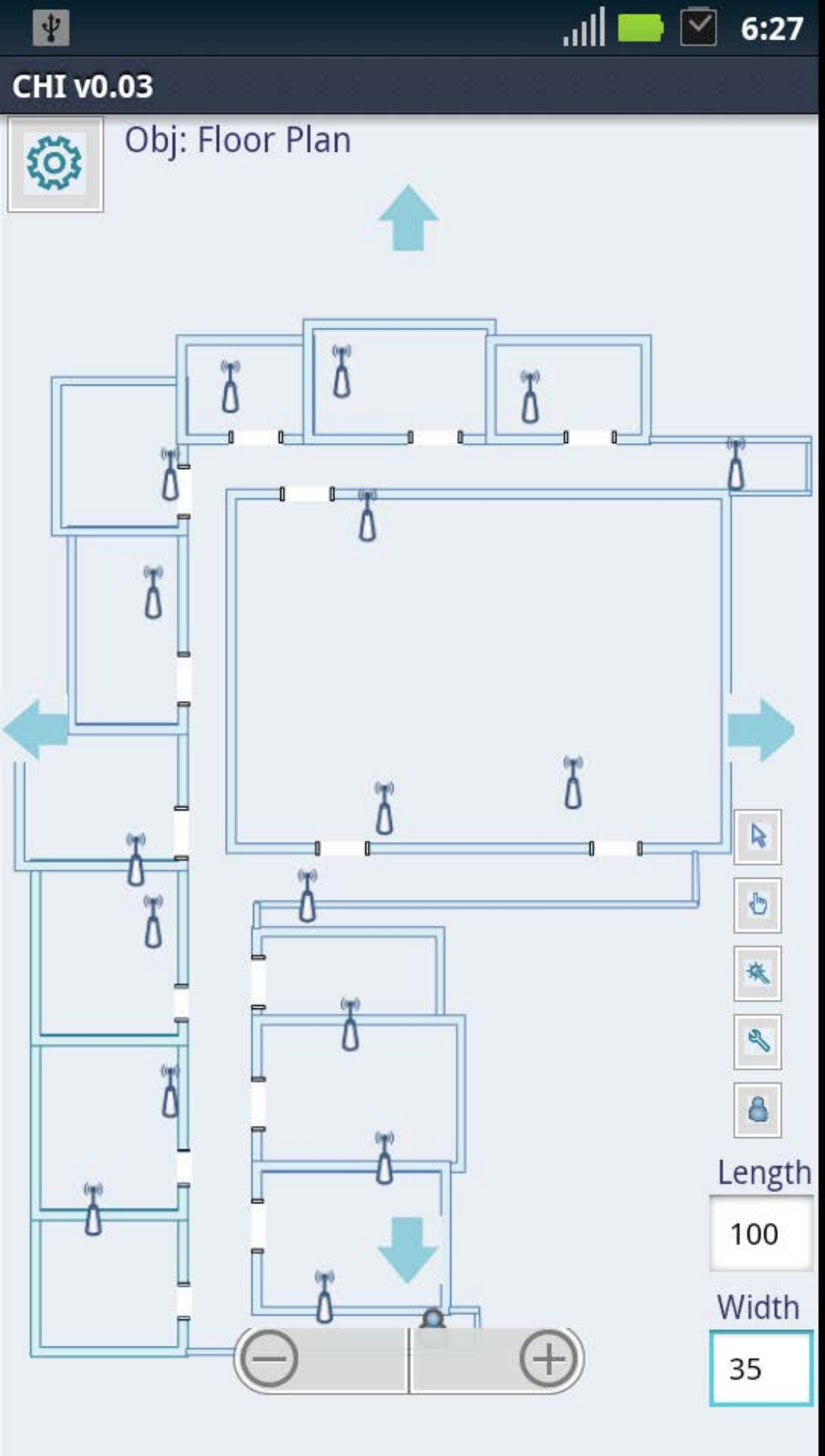}
}
\hspace{0.05in}
\subfigure[The result corrected by the laborer] { \label{fig6:f}
\centering
\includegraphics[width=1.5in,height=2.8in]{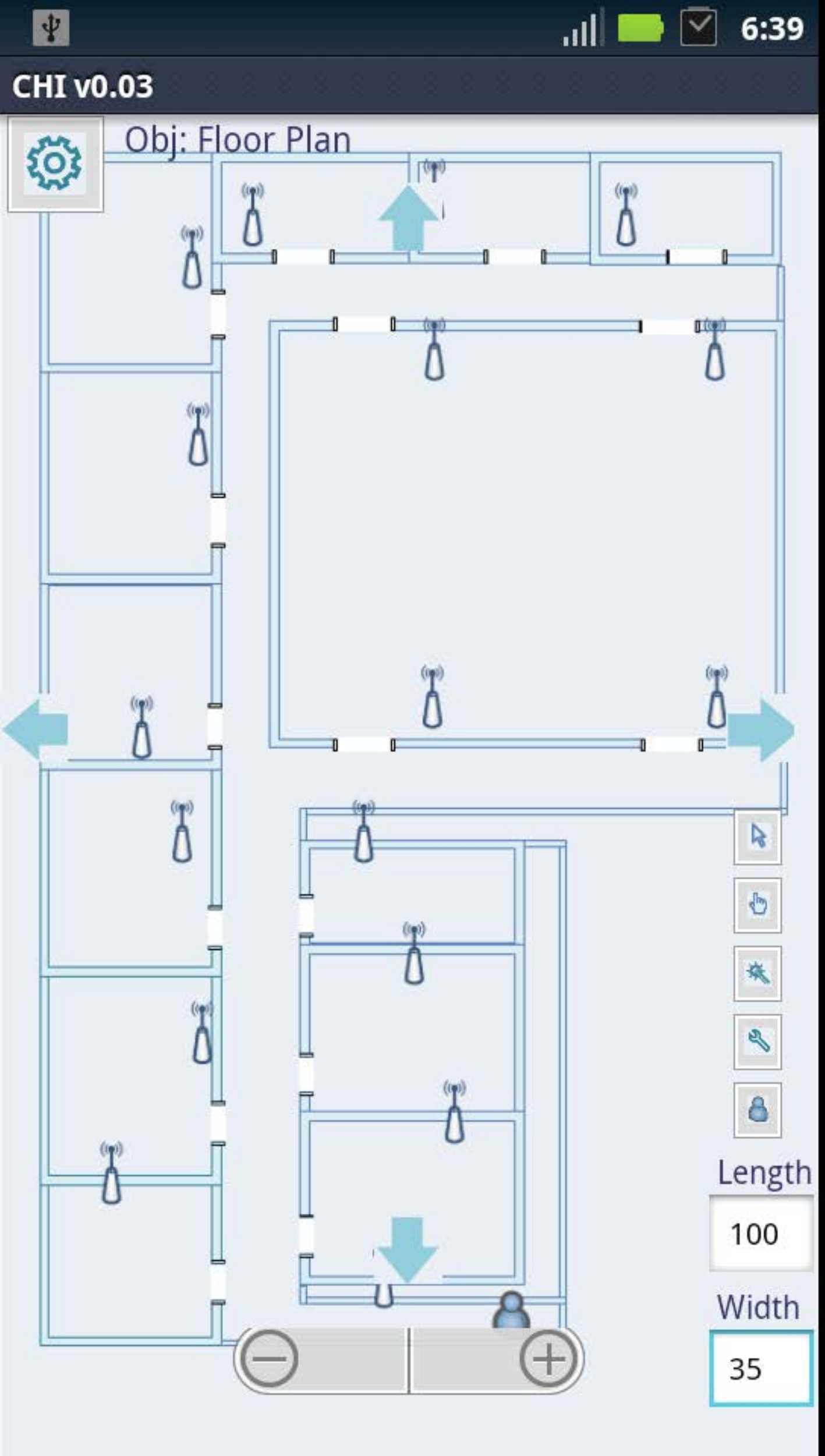}
}
\hspace{0.05in}
\subfigure[The status of the AP-to-AP trajectories] { \label{fig6:g}
\centering
\includegraphics[width=1.5in,height=2.8in]{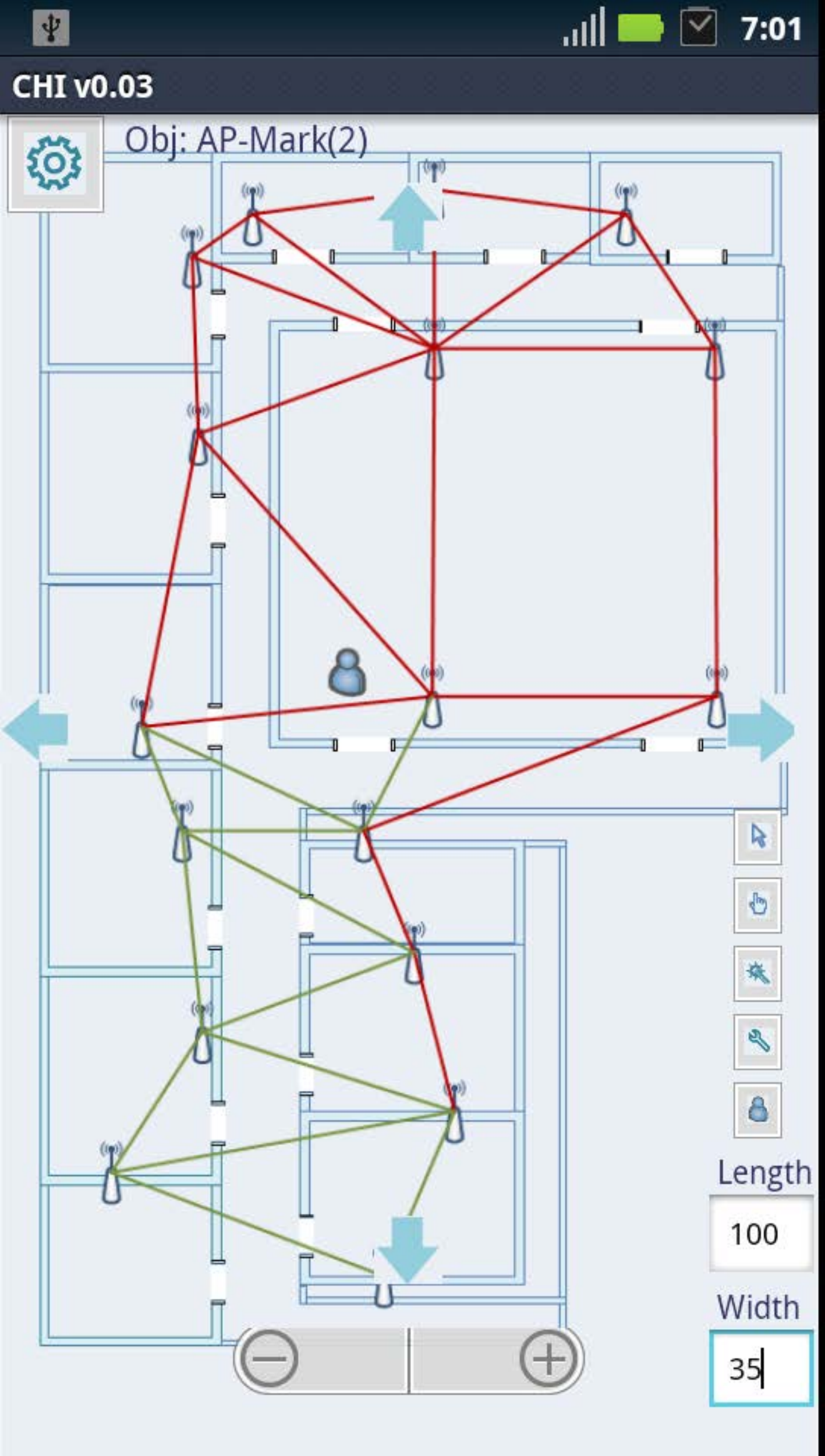}
}
\hspace{0.05in}
\subfigure[The trajectory of the laborer during a period of time] { \label{fig6:h}
\centering
\includegraphics[width=1.5in,height=2.8in]{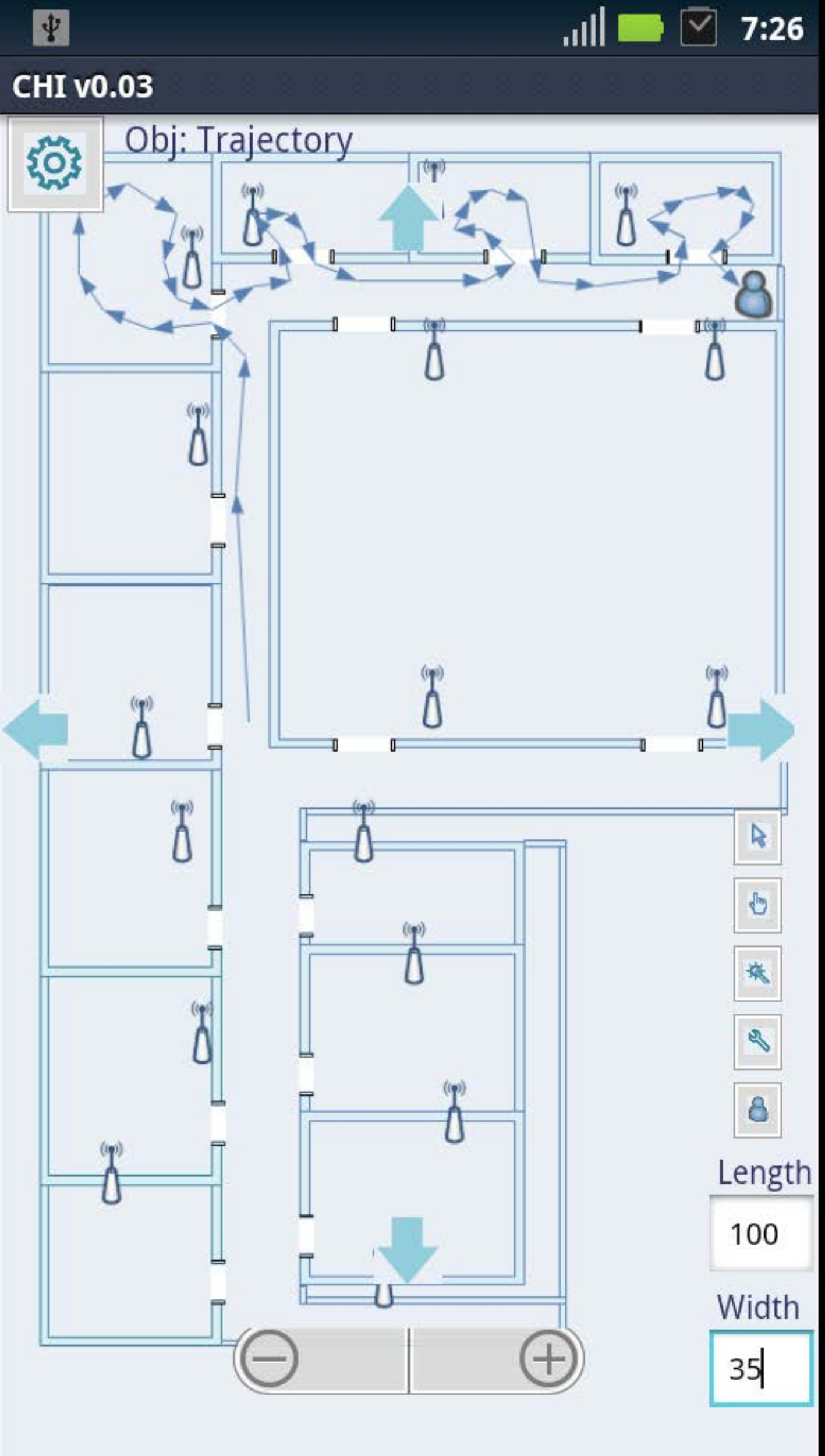}
}
\caption{Example of location-based services}
\label{fig6}
\end{figure*}

\section{Use Case of CHI} \label{use_case}
We describe an use case of CHI in this section. The floor plan with respect to the use case is shown in Figure \ref{fig5}\footnote{The access point $AP_4$, $AP_5$, $AP_{10}$, $AP_{13}$, $AP_{15}$ and $AP_{17}$ are provided by mobile phones, and other access points are wireless routers.}. In the use case, the laborer firstly selects the sub-service 1 of locating the APs of interest (Section \ref{sec_subs1}) in CHI, and targets at exploring all possible APs in the floor area at the beginning. As a feedback, CHI suggests the pathway to the laborer as shown in Figure \ref{fig6:a}. Then, with the laborer's walking, CHI would continuously generates its results on the area that has been explored. At the time of three rooms having been explored, the laborer stops to correct the size of these three rooms and the location of three found APs generated by CHI, and lock his correction result. (His correction result is shown in Figure \ref{fig6:b}.) With these three rooms and three APs locked by the laborer, CHI would not modify them again. Next, the laborer decides to change and re-order his objectives in CHI. As shown in Figure \ref{fig6:c}, the laborer set the list of objectives to \texttt{<Floor plan, AP-Mark(1)>}. On receiving the new list of objectives, CHI would change to regard the floor plan as its primary objective, and provide the pathway suggestion of the floor plan (introduced in Section \ref{sec_floorplan}) as shown in Figure \ref{fig6:d}. Compared to the pathway suggestion of locating the APs, the pathway suggestion of the floor plan required finer grained sampling points. Following the pathway suggestion, the laborer continues to walk in the floor area. And after a time period, when the laborer finishes his walking with respect to the floor plan, CHI gives a raw result of the floor plan as shown in Figure \ref{fig6:e}. Then, the laborer corrects this raw result and locks the corrected floor plan. (The corrected floor plan is shown in \ref{fig6:f}.) At the same time, CHI would check to find that both the two objectives in list \texttt{<Floor plan, AP-Mark(1)>} have been completed and remove them. With all objectives having been completed, this use case is finished.

In addition, the laborer may want to improve the the quality of the AP-to-AP trajectories, and select the sub-service 2 of locating the APs of interest (Section \ref{sec_subs2}) in CHI. Then, after the selection, CHI would exhibit the status of the AP-to-AP trajectories to the laborer with red line indicating "the trajectory needs to be improved" and green line indicating "the trajectory has a good quality". Figure \ref{fig6:g} shows the status of the AP-to-AP trajectories at a time during the laborer's walking. Besides, the laborer could select to track his trajectory in CHI. Figure \ref{fig6:h} shows the display result of the laborer's trajectory during a certain period of time.

\section{Evaluation} \label{evaluation}
We compare the CHI-based approach with the Fingerprinting-based approach\footnote{Note that the Fingerprinting-based approach applied in our evaluation scenario has some differences with the finerprinting-based approach understood in the usual sense. The Fingerprinting-based approach in our scenario fingerprints at the location of each AP for its displacement with neighbor APs, while the finerprinting-based approach usually understood collects the RSS of signals from nearby APs at the sampling locations that may not be the location of APs.} and the Crowdsourcing-based approach on time cost and expense cost by simulations and the experiment on practical data with respect to the use case of CHI presented in last Section. In simulations, we randomly deploy $100$ APs in a $100\times100$ square area, and divide the area around each AP into $8$ equal sectors. If an AP has neighbor APs in one of its sectors, we choose to build or not to build an AP-to-AP trajectory with the probability of $0.5$ between this AP and its nearest neighbor AP in this sector. After the building process, if there still exist isolated APs in the square area, we establish an AP-to-AP trajectory between each isolated AP and its nearest neighbor AP. Suppose that both the CHI-based approach and the Crowdsourcing-based approach measures the direction between APs within the error of $\pm 30$ degrees and the length between APs within the error of $\pm 10$ percent, and the Fingerprinting-based approach measures the direction and length within $\pm 30p$ degrees and $\pm 10p$ percent, $0< p <1$, respectively. Although the Fingerprinting-based approach has higher accuracy than the CHI-based approach and the Crowdsourcing-based approach in measurement, it often requires the laborer to spend more time to obtain the measurement result. Thus, we assume that the laborer takes $1$ time unit per length to obtain the measurement result for the CHI-based approach and the Crowdsourcing-based approach, and takes $c$ time unit per length for the Fingerprinting-based approach, $c>1$.

In evaluation, CHI-based approach apply the method presented in Section \ref{sec_subs1} to build the shortest Hamilton path (with its start location on left bottom of the area and its start direction from left to right) for pathway suggestion, and the laborer walks to collect and measure the AP-to-AP trajectories for each AP following the sequence of APs on this Hamilton path. For the Fingerprinting-based approach, we assume that the laborer applies the same pathway walking as the CHI-based approach, and for Crowdsourcing-based approach, assume the laborer collects the AP-to-AP trajectories by random walk with a random start location.

Next, we study the relationship between the localization accuracy of APs' relative location and the time/expanse cost for these three approaches since the localization accuracy of APs' relative location would determine the quality of location-based services presented in Section \ref{sec_lbs}. All these three approaches applies the Arturia algorithm \cite{Walkie-markie} for localization. The location of all APs are assigned to the coordinate of $(0,0)$ at initialization time, and the running process of Arturia algorithm would terminate after $100$ iterations.

\begin{figure*}[!htb] \centering
\subfigure[Ground truth] { \label{ef1:a}
\raggedleft
\includegraphics[width=1.58in]{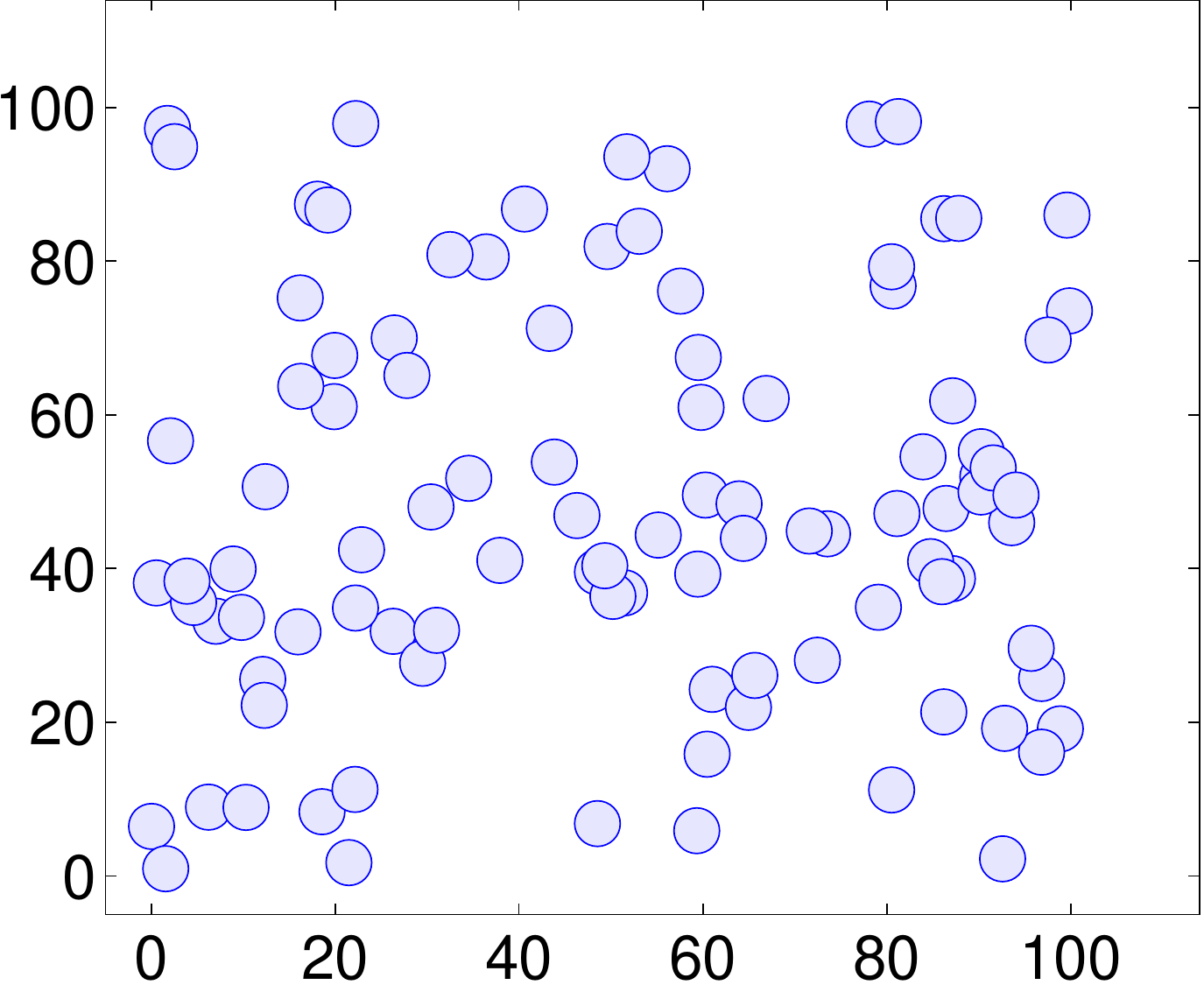}
}
\hspace{0.05in}
\subfigure[CHI-based approach] { \label{ef1:b}
\raggedleft
\includegraphics[width=1.58in]{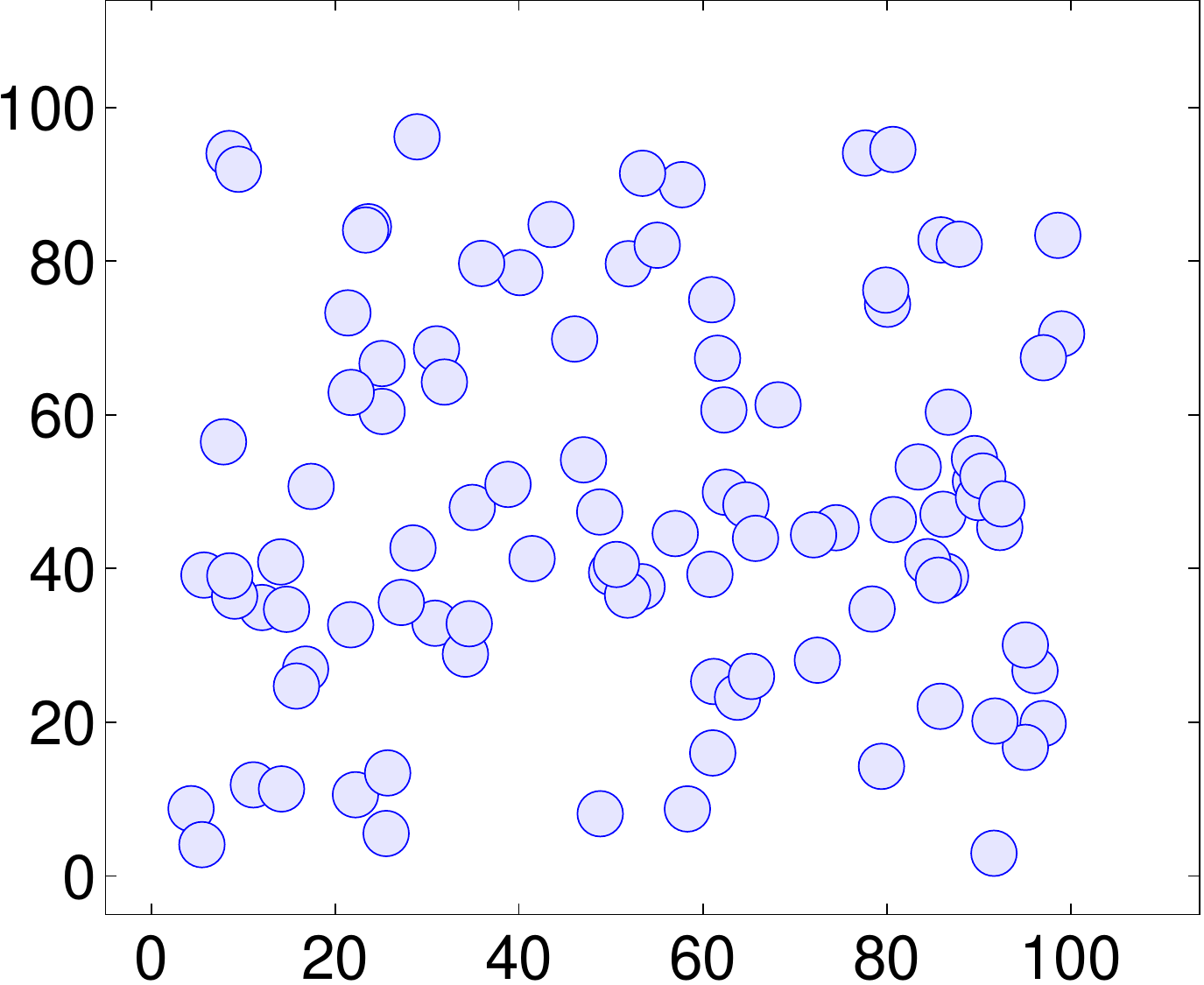}
}
\hspace{0.05in}
\subfigure[Crowdsourcing-based approach with $\#crowds=5$] { \label{ef1:c}
%\centering
\raggedleft
\includegraphics[width=1.58in]{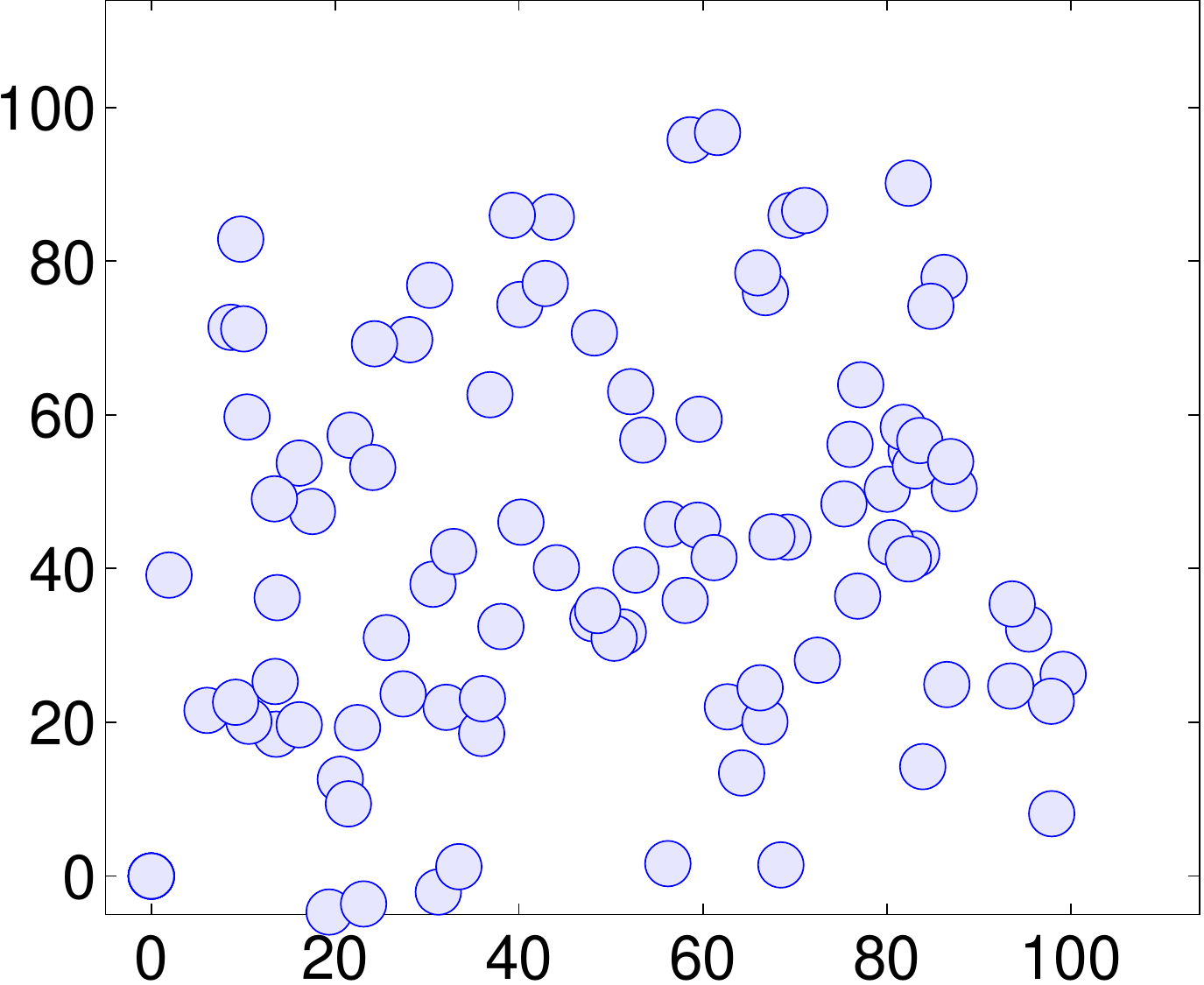}
}
\hspace{0.05in}
\subfigure[Fingerprinting-based approach with $p=\frac{1}{5}$ and $c=5$] { \label{ef1:d}
\raggedleft
\includegraphics[width=1.58in]{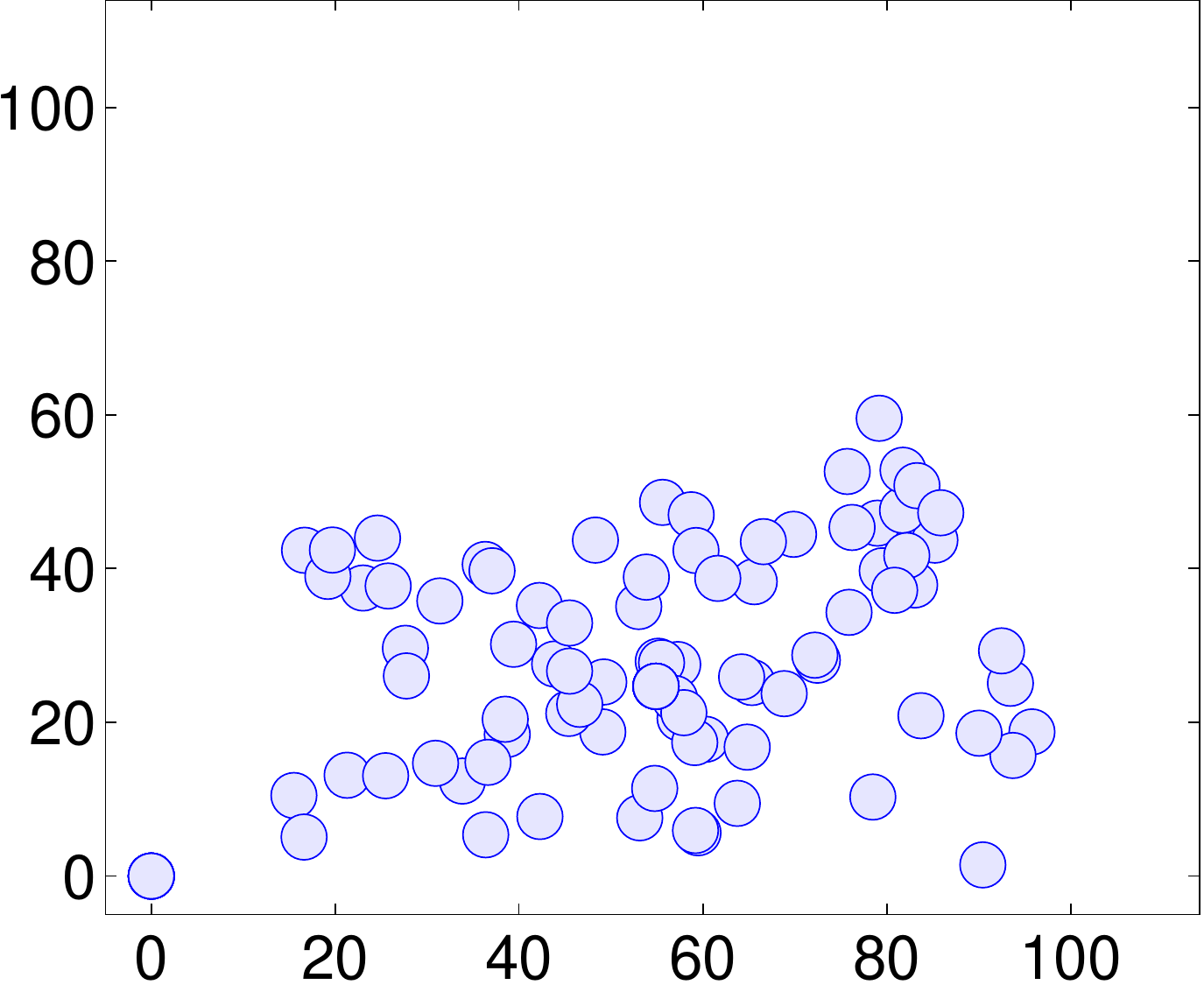}
}
\caption{The snapshot of localization result at $8000$ time units}
\label{ef1}
\end{figure*}

\begin{figure*}
\begin{minipage}[]{0.24\textwidth}
\centering\includegraphics[height=1.50in,width=1.65in]{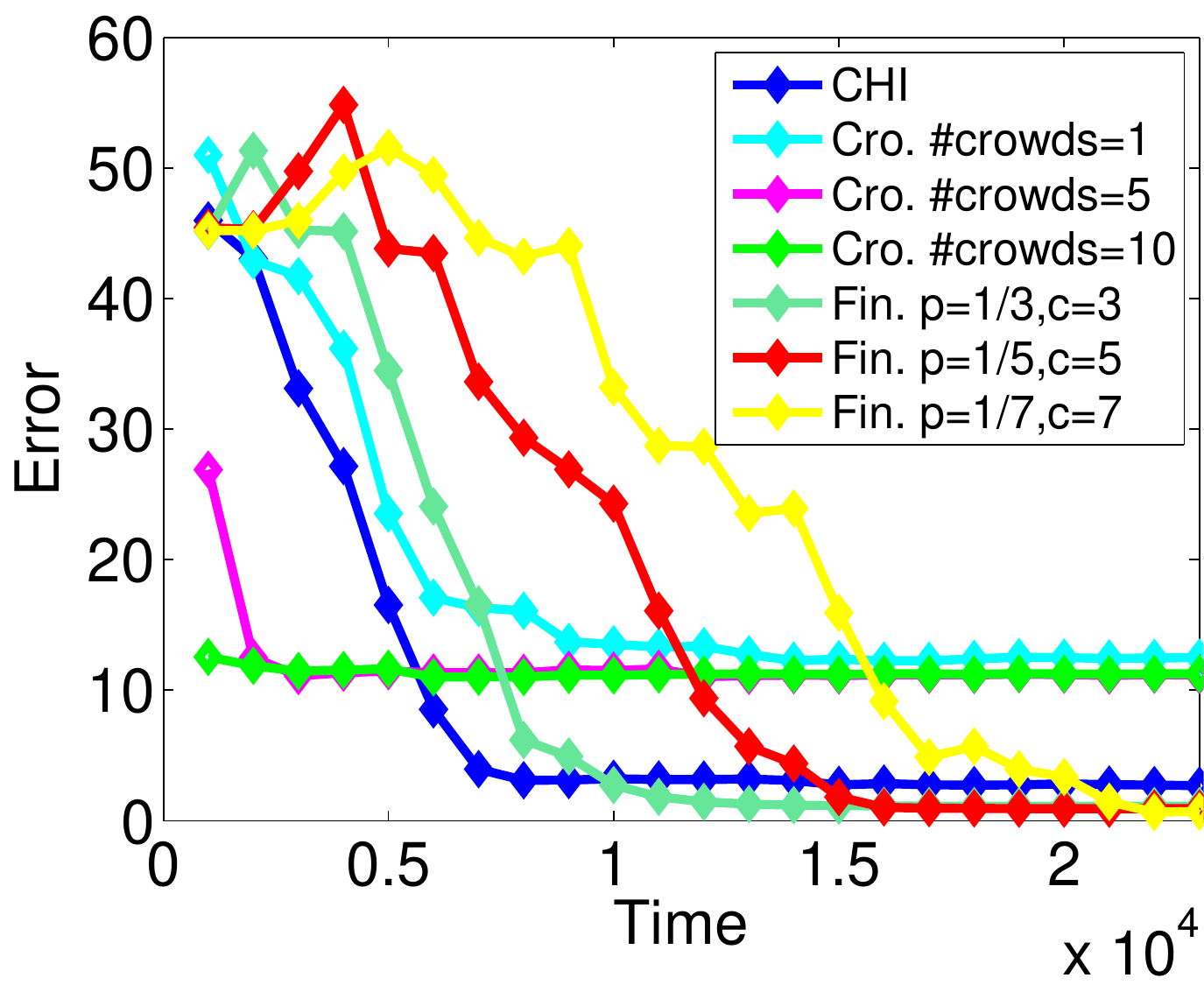}
\caption{\textrm{The time cost for the $100 \times 100$ square area}}\label{ef2}
\end{minipage}
\begin{minipage}[]{0.24\textwidth}
\centering\includegraphics[height=1.43in,width=1.65in]{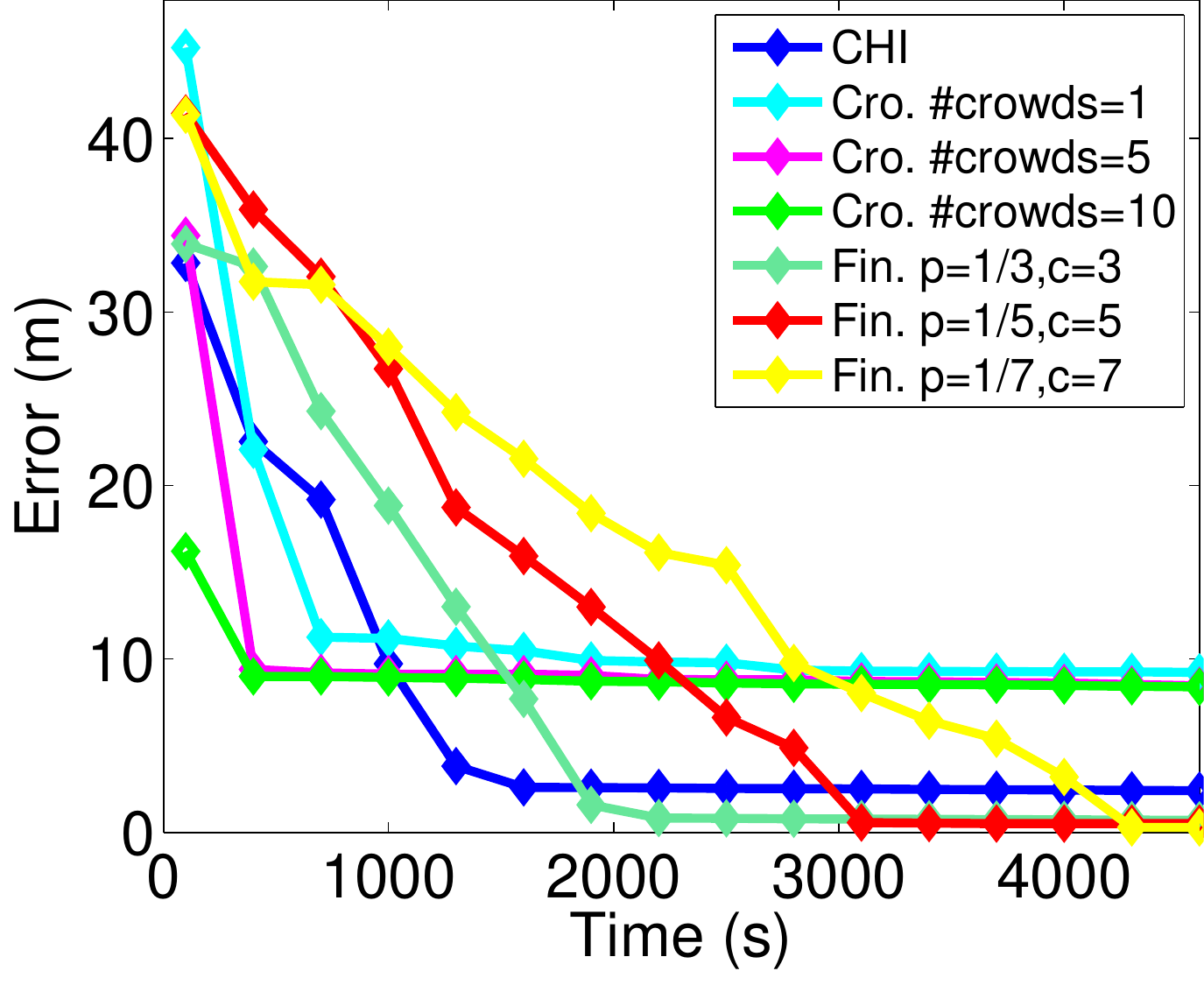}
\caption{\textrm{The time cost for the use case in Section \ref{use_case}}}\label{ef3}
\end{minipage}
\begin{minipage}[]{0.25\textwidth}
\centering\includegraphics[height=1.40in,width=1.72in]{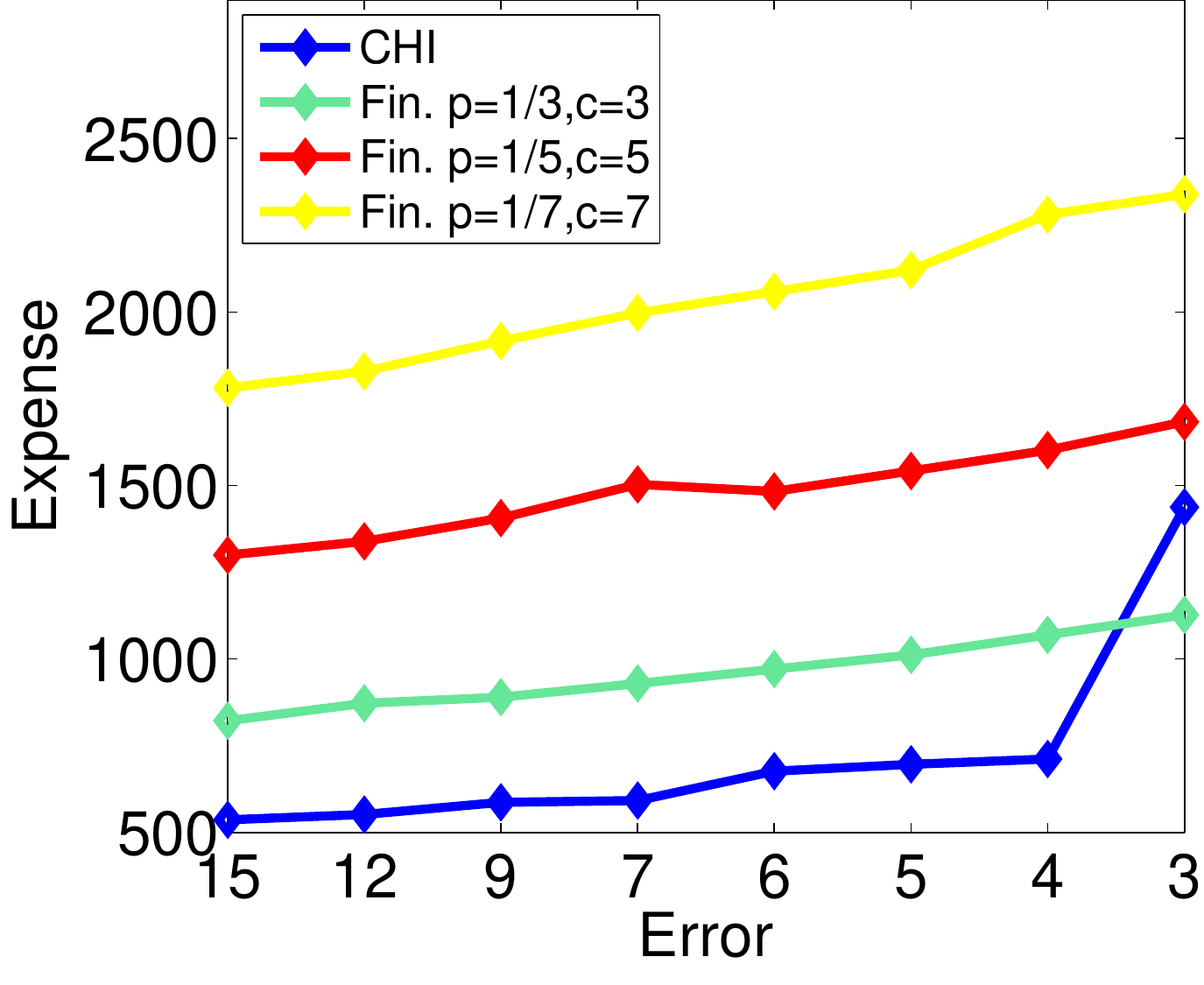}
\caption{\textrm{The expense cost for the $100 \times 100$ square area}}\label{ef4}
\end{minipage}
\begin{minipage}[]{0.25\textwidth}
\centering\includegraphics[height=1.40in,width=1.72in]{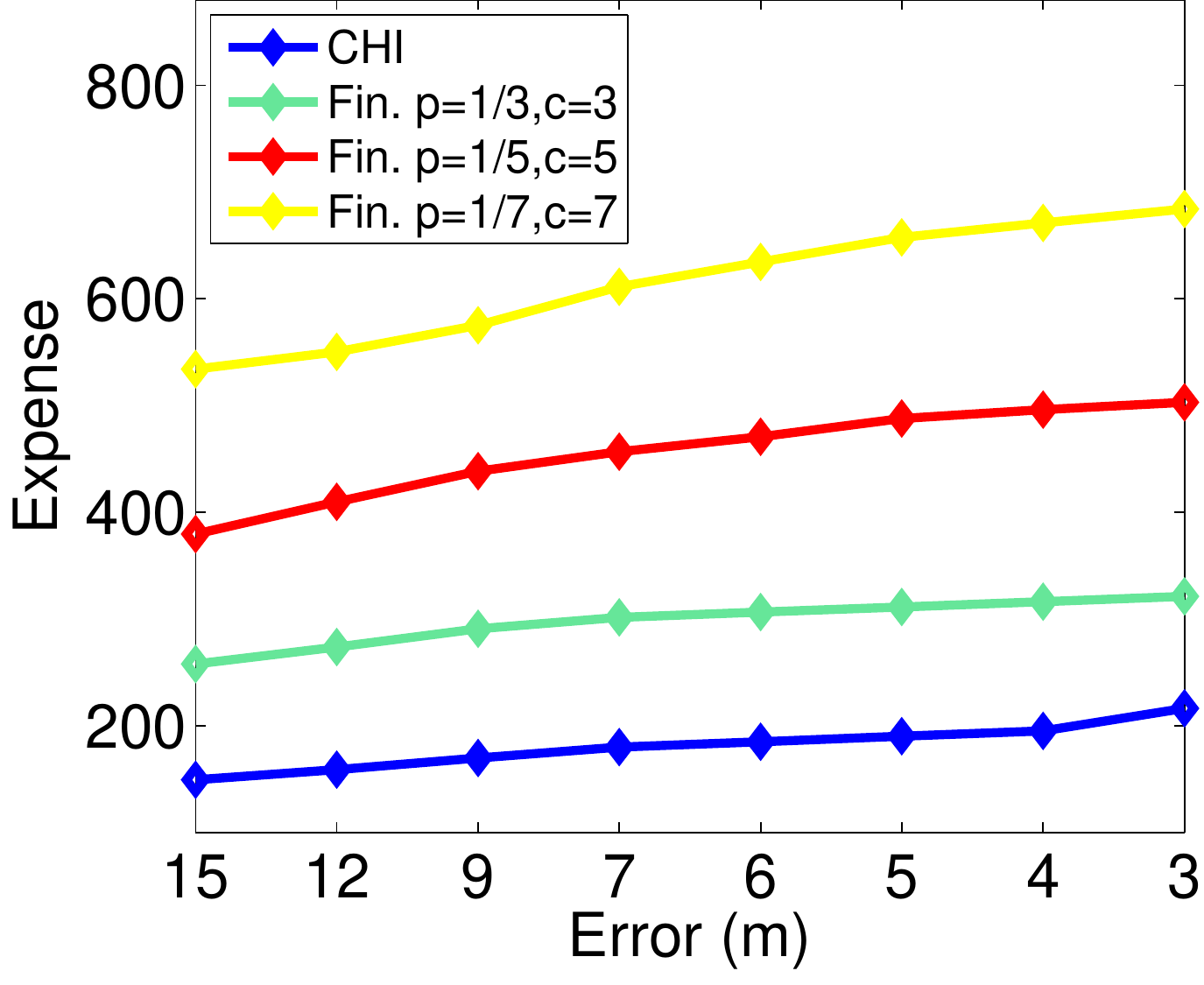}
\caption{\textrm{The expense cost for the use case in Section \ref{use_case}}}\label{ef5}
\end{minipage}
\end{figure*}

\subsection{Time Cost}
Compared with the other two approaches, the localization process applying the CHI-based approach (called the CHI-based process for short) has the fastest convergence rate in reducing the localization errors. Figure \ref{ef1} gives the snapshot of localization result on APs' location at $8000$ time units for the localization process, CHI, crowdsourcing with $\#crowds=5$, and fingerprinting with $p=\frac{1}{5}$ and $c=5$, respectively. As can be seen, the localization results of the CHI-based process is quite close to the ground truth, having the average localization error of $2.86$ (length units) for the location of APs, while the localization result of the Crowdsourcing-based process and the Fingerprinting-based process have a far more larger error of $11.17$ and $29.09$ respectively.

Further, we set the parameter $(p,c)$ to $(\frac{1}{3},3)$, $(\frac{1}{5},5)$ and $(\frac{1}{7},7)$ respectively for evaluation of the Fingerprinting-based approach. The setting of $(p,c)$ is based on the fact that the laborer with the same measurement device would expend more effort to obtain the measurement result of less error, e.g., spending $5$ time units to obtain $\frac{1}{5}$ measurement error in length vs. spending $7$ time units to obtain $\frac{1}{7}$ measurement error. We set that there are $1$, $5$ and $10$ laborers respectively working for evaluation of the Crowdsourcing-based approach. Figure \ref{ef2} shows the evaluation result of these approaches on average localization error of APs over time. As can be seen, (1) the localization error of the CHI-based process falls with the fastest rate over the time period of $0-8000$ time units, and keeps around $2.7$ after $8000$ time units; (2) the localization error of the Crowdsourcing-based process has a relative low value initially for $\#crowds=5$ and $\#crowds=10$ with the advantage on the number of laborers working in parallel, but keeps more than $11.0$ soon since random walk would miss some APs that are hard to visit with few neighbor APs; (3) Employing more laborers has a little effect in reducing the localization error because the activity trajectories of the laborers focus on the central APs that have more neighbor, and still ignore the APs with less neighbors; (4) the localization of the Fingerprinting-based process is quite large at the beginning, but reduce speedy to even less than the error of the CHI-based process after a period of fingerprinting time since the advantage of more accurate measurement would appear once the laborer completes or nearly completes the fingerprinting; (5) more effort in reducing the measurement error would increase the time cost, but make the localization process getting more accurate, e.g., the Fingerprinting-based process having average localization error of around $1.12$ after $16000$ time units for $p=\frac{1}{3}$ and $c=3$, and around $0.69$ after $22000$ time units for $p=\frac{1}{7}$ and $c=7$.

For the practical use case in Section \ref{use_case}, the laborer preliminarily collects a plenty of AP-to-AP trajectories by $1$ hour walking in the floor area. The CHI-based process and Fingerprinting-based process select the trajectories by the sequence of the APs on the shortest Hamilton path, and the Crowdsourcing-based process select the trajectories by random walk with its start point at the location of a randomly selected AP. We get the similar result as in simulation. As shown in Figure \ref{ef3}, by applying the CHI-based process, the average localization error on APs' location reduces with the fastest rate over the time period of $0-1600$ seconds and keeps around $2.3$ meters after $1600$ seconds, while the localization error of the Crowdsourcing-based process keeps more than $8$ meters all the time, and the localization error of the Fingerprinting-based process reduces to lower than $1$m with more accurate measurement and more time spent.

\subsection{Expense Cost}
We use the following equation for comparing the expense cost of the approaches:
\begin{equation} \label{e3}
E=t\cdot e_l + b \cdot e_d
\end{equation}
Where $e_l$ is the expense on employing a labor per time unit, and $t$ is the total working time of all laborers; $e_d$ is the expense on renting or buying the measurement device, and $b$ is the number of measurement devices.

Figure \ref{ef4} gives a result on expense cost by choosing the parameter $(e_l(CHI),e_d(CHI))=(0.1,36)$, $(e_l(Fin.),e_d(Fin.))=(0.1,\frac{36}{p})$ for the Fingerprinting-based approach, $e_l(Cro.)\approx 0$ and $e_d(Cro.) = 0$ for the Crowdsourcing-based approach. In this parameter setting, we set $e_d(Fin.)>e_d(CHI)$ based on the common experience that the device with more accurate measurement has a higher expense cost, and set $e_l(Cro.)\approx 0 \ll e_l(CHI), e_l(Cro.)$, $e_d(Cro.) = 0$, that is to say, the expense cost of crowdsourcing is negligible, since the crowds unconsciously join to collect the data of trajectories with their own devices and by their normal activities in the floor area. The average localization error of APs varies on $<15,12,9,7,6,5,4,3>$, and the result in Figure \ref{ef4} shows that (1) applying CHI-based approach has lower expense cost than applying the Fingerprinting-based approach; (2) more accurate approach needs more expense cost, e.g., Fin. $p=\frac{1}{7}$ and $c=7$ vs. Fin. $p=\frac{1}{5}$ and $c=5$; (3) for the case of average localization error of $3$, the process applying the CHI-based approach has a higher expense cost than applying fingerprinting-based approach(, and no matter how to increase expense cost, the average localization error of CHI-based approach keeps around $2.7$ finally), since CHI-based measurement has a bottleneck on accuracy and is hard to improve the localization accuracy after the average localization error reduced to a certain value. The expense cost result of the use case of CHI is shown in Figure \ref{ef5}. This result applies the same parameter setting as in simulation, and we can infer similar conclusions like from Figure \ref{ef4} (except for the average error of $3$m case, since there are not too much APs in the use case). Also, we try to vary the parameter $e_l$ and $e_d$, and get the similar results\footnote{The detail results is omitted here due to space limit.}. Especially, when increasing $e_d$, applying the CHI-based approach has a even lower expense than applying the Fingerprinting-based approach.

\subsection{Discussion}
The localization process applying the Crowdsourcing-based approach has relatively larger error than the process applying the other two approaches in both the $100 \times 100$ square case and the practical use case as shown in evaluation results. The reason is that the random walk strategy in the Crowdsourcing-based approach would miss the APs located in the forgotten area of the floor. The Crowdsourcing-based approach, as its name implies, applies to the scenario that the area of interest is frequently visited by the crowd, and also the implementer should persuade the crowd to use the application and allow the application to upload data to the back-end server.

The localization process applying the fingerprinting-based approach could achieve lower error than applying the other two approaches by more accurate measurement, but requires more time cost and expense cost as shown in evaluation results. The Fingerprinting-based approach applies to the scenario that the demander has the need of reducing or controlling the localization error and could tolerate the time and expense cost.

The CHI-based approach considers both the Crowdsourcing-based fashion that collects and measures the trajectory of the laborer and the Fingerprinting-based fashion that plans to achieve the localization objectives, and is implemented without any assumption of priori knowledge on the floor area. A single user can plans to achieve a set of localization objectives just by walking through the area of interest.

\section{Related Works} \label{related}
CHI-based system is related to the work on multiple fields of localization, chiefly RSS-based localization, IMU-based tracking, and Simultaneous Localization and Mapping (SLAM) problem in robotics fields.

Localization on utilizing the RSS of signal has been intensively studied for many years. A popular approach is the Fingerprinting-based approach that establishes the mapping of RSS-to-location for localization. Radar \cite{Radar} is the first proposed method on Fingerprinting-based localization that estimates the location of the target by the knowledge of APs' location, transmit power and the floor map. Horus \cite{Horus} built a probability distribution in RSS-to-location mapping and located the target by matching the mapping with maximum likelihood. A variety of improvements or extensions have also been made on Fingerprinting-based localization, such as considering the mobility constraints \cite{Mobility} and incorporating ambient signatures from environment \cite{SurroundSense}. Another approach on utilizing the RSS is the Model-based approach that directly build the relationship between RSS and distance e.g. the log-distance path loss model \cite{Log-distance}. Efforts like locating by APs \cite{Model_AP} and learning the ability of measurement \cite{Model_ability} have been made based on the model of RSS-to-distance.

The RSS-based localization is usually incorporated into the IMU-based tracking, since tracking the trajectory of the user merely by IMU sensors equipped in mobile phone can suffer from accumulate error over time \cite{IMU_error}. Unloc \cite{Unloc} calibrated the IMU-based tracking by various types of landmarks, including the WiFi landmarks built on the similarity metric of the ratio of common APs. Pedestrian-tracking \cite{Pedestrian_tracking} trained to obtain maximum RSS points and used them as switches between location inference modules. Walkie-Markie \cite{Walkie-markie} tracked the pathway of the user by WiFi-Marks, which are the maximum RSS points in directions built on the data collected from the crowd. Furthermore, the IMU-based tracking of the crowds (or the Crowdsourcing-based approach) could help reducing the effort made for the RSS-based localization, e.g., LiFS \cite{LiFS} and Zee \cite{Zee} sought to reduce the effort of building the RSS-to-location mapping by the trajectories of the crowd.

There has been much work on dealing with the SLAM problem in robotics fields. Usually, in the scenario of SLAM, a robot is designated to explore the area of interest for simultaneous localization and mapping with equipped sensors (e.g. camera, laser and sonar). The localization problem is to determine the relative location of the robot in the area, and the mapping problem is to locate the objects surrounding the robot. For example, FootSLAM \cite{FootSLAM} built the floor plan with shoe-mounted IMU sensors, and WiFi-SLAM \cite{WiFi-SLAM} used the Gaussian process latent variable model to build RSS-based connectivity graphs.

\section{Conclusions} \label{conclusion}
We study the CHI-based approach by implementing a system on mobile phone. The laborer can set a list of localization objectives in CHI, and CHI provides the pathway suggestion and the solution of these objectives for the laborer based on its trajectory management, which applies AP-based trajectory calibration to improve the quality of the collected trajectories. In evaluation, we compare the CHI-based approach with two other approaches and show its advantage in reducing the time cost and saving the expense cost with respect to the achievement on average localization error of APs.


\begin{thebibliography}{99}

\vspace{-0.3ex}

\bibitem{SurroundSense}
M. Azizyan, I. Constandache, and R. Roy Choudhury. SurroundSense: Mobile Phone Localization via Ambience Fingerprinting. In MobiCom, 2009.

\vspace{-0.3ex}

\bibitem{Radar}
P. Bahl and V. N. Padmanabhan. RADAR: An Inbuilding RF-based User Location and Tracking System. In INFOCOM, 2000.

\vspace{-0.3ex}

\bibitem{Kalman}
B. Barshan and H. Durrant-Whyte. Inertial navigation systems for mobile robots. In IEEE Trans. on Robotics and Automation, Vol. 11, No. 3, 1995.

\vspace{-0.3ex}

\bibitem{IMU_error}
S. Beauregard and H. Haas. Pedestrian Dead Reckoning: A Basis for Personal Positioning. In WPNC, 2006.

\vspace{-0.3ex}

\bibitem{Step_comparison}
A. Brajdic, and R. Harle. Walk Detection and Step Counting on Unconstrained Smartphones. In UbiComp, 2013.

\vspace{-0.3ex}

\bibitem{Step_length}
D. K. Cho, M. Mun, U. Lee, W. J. Kaiser, and M. Gerla. Autogait: A mobile platform that accurately estimates the distance walked. In PerCom, 2010.

\vspace{-0.3ex}

\bibitem{Escort}
I. Constandache, X. Bao, M. Azizyan, and R. R. Choudhury. Did you see bob?: human localization using mobile phones. In MobiCom, 2010.

\vspace{-0.3ex}

\bibitem{WiFi-SLAM}
B. Ferris, D. Fox, and N. Lawrence. Wifi-slam using gaussian process latent variable models. In IJCAI, 2007.

\vspace{-0.3ex}

\bibitem{Model_ability}
A. Goswami, L. E. Ortiz, and S. R. Das. WiGEM : A Learning-Based Approach for Indoor Localization. In CoNEXT, 2011.

\vspace{-0.3ex}

\bibitem{PI}
Z. Guo, Y. Guo, F. Hong, X. Yang, Y.He, Y. Feng, and Y. Liu. Perpendicular Intersection: Locating Wireless Sensors with Mobile Beacon. In RTSS, 2008.

\vspace{-0.3ex}

\bibitem{Mobility}
A. Haeberlen, E. Flannery, A. M. Ladd, A. Rudys, D. S. Wallach, and L. E. Kavraki. Practical Robust Localization over Large-Scale 802.11 Wireless Networks. In MobiCom, 2004.

\vspace{-0.3ex}

\bibitem{Pedestrian_tracking}
Y. Kim, H. Shin, and H. Cha. Smartphone-based wi-fi pedestriantracking system tolerating the rss variance problem. In PerCom, 2012.

\vspace{-0.3ex}

\bibitem{Model_AP}
H. Lim, L. Kung, J. Hou, and H. Luo. Zero-Configuration, Robust Indoor Localization: Theory and Experimentation. In INFOCOM, 2006.

\vspace{-0.3ex}

\bibitem{Zee}
A. Rai, K. K. Chintalapudi, V. N. Padmanabhan, and R. Sen. Zee: zero-effort crowdsourcing for indoor localization. In MobiCom, 2012.

\vspace{-0.3ex}

\bibitem{Log-distance}
T. Rappaport. Wireless Communications Principles and Practice. Prentice Hall, 2001.

\vspace{-0.3ex}

\bibitem{FootSLAM}
P. Robertson, M. Angermann, and B. Krach. Simultaneous localization and mapping for pedestrians using only foot-mounted inertial sensors. In UbiComp, 2009.

\vspace{-0.3ex}

\bibitem{MCD}
P. Rousseeuw. Multivariate Estimation with High Breakdown Point. In Mathematical Statistics and Applications, vol. 8, 1985.

\vspace{-0.3ex}

\bibitem{Fast_MCD}
P. Rousseeuw and K. V. Driessen. A Fast Algorithm for the Minimum Covariance Determinant Estimator. In Technometrics, vol. 41, no. 3, 1999.

\vspace{-0.3ex}

\bibitem{Walkie-markie}
G. Shen, Z. Chen, P. Zhang, T. Moscibroda, and Y. Zhang. Walkie-markie: Indoor pathway mapping made easy. In NSDI, 2013.

\vspace{-0.3ex}

\bibitem{Unloc}
H. Wang, S. Sen, A. Elgohary, M. Farid, M. Youssef, and R. R. Choudhury. No Need to War-Drive: Unsupervised Indoor Localization. In Mobisys, 2012.

\vspace{-0.3ex}

\bibitem{LiFS}
Z. Yang, C. Wu, and Y. Liu. Locating in fingerprint space: wireless indoor localization with little human intervention. In MobiCom, 2012.

\vspace{-0.3ex}

\bibitem{Horus}
M. Youssef and A. Agrawala. The Horus WLAN Location Determination System. In MobiSys, 2005.

\vspace{-0.3ex}

\bibitem{Robust_trajectory}
X. Zhang, Z. Yang, C. Wu, W. Sun, Y. Liu, and K. Liu: Robust Trajectory Estimation for Crowdsourcing-Based Mobile Applications. In IEEE Trans. on Parallel and Distributed Computing, Vol. 25, No. 7, 2014.

\end{thebibliography}
\end{document}